\documentclass[11pt]{article}
\newenvironment{wileykeywords}{\textsf{Keywords:}\hspace{\stretch{1}}}{\hspace{\stretch{1}}\rule{1ex}{1ex}}

\usepackage{amsmath,amssymb}
\usepackage{graphicx}
\usepackage{color}
\usepackage{dcolumn}
\usepackage{bm}
\usepackage[numbers,super,comma,sort&compress]{natbib}

\definecolor{background-color}{gray}{0.98}

\newcommand{\rmi}{{\rm i}}
\newcommand{\rme}{{\rm e}}
\newcommand{\rmd}{{\rm d}}

\def\bra#1{\mbox{$\langle#1|$}}
\def\ket#1{\mbox{$|#1\rangle$}}

\title{Efficient calculation of open quantum system dynamics and time-resolved spectroscopy with Distributed Memory HEOM (DM-HEOM).}

\author{Tobias Kramer\thanks{Zuse Institute Berlin (ZIB), Takustr.\ 7, 14195 Berlin, Germany}
\thanks{Department of Physics, Harvard University, 17 Oxford Street, 02138 Cambridge, Massachusetts, USA} 
\and Matthias Noack\footnotemark[1]
\and Alexander Reinefeld\footnotemark[1] 
\and Mirta Rodr\'iguez\footnotemark[1]
\and Yaroslav Zelinskyy\footnotemark[1]}

\begin{document}

\maketitle

\begin{abstract}
Time- and frequency resolved optical signals provide insights into the
properties of light harvesting molecular complexes, including excitation
energies, dipole strengths and orientations, as well as in the exciton
energy flow through the complex.

The hierarchical equations of motion (HEOM) provide a unifying theory,
which allows one to study the combined effects of system-environment
dissipation and non-Markovian memory without making restrictive assumptions about
weak or strong couplings or separability of vibrational and electronic
degrees of freedom.

With increasing system size the exact solution of the open quantum system dynamics
requires memory and compute resources beyond a single compute node.
To overcome this barrier, we developed a scalable variant of HEOM.
Our distributed memory HEOM, DM-HEOM, is a universal tool for open quantum system dynamics.
It is used to accurately compute
all experimentally accessible time- and frequency resolved processes  in
light harvesting molecular complexes with arbitrary system-environment
couplings for a wide range of temperatures and complex sizes.
\end{abstract}

\begin{wileykeywords}
time-dependent spectroscopy; light-harvesting complexes; numerical methods; open quantum systems; non-Markovian environment
\end{wileykeywords}

  \makeatletter
  \renewcommand\@biblabel[1]{#1.}
  \makeatother

\bibliographystyle{apsrev}

\section*{\sffamily \Large INTRODUCTION}

Quantum systems at finite temperatures are not perfectly shielded from the surrounding matter and exchange energy with their environment.
This has diametrical effects on the quantum-mechanical coherence of the system and is the main obstacle for engineering usable quantum information processing gates and Qbits.
The environment is typically considered as many vibrational states (the ``bath'') kept at fixed temperature, while the ``system'' is brought to a non-equilibrium state by an external time-dependent perturbation.
In molecular complexes, the environment drives electronic excitations towards their equilibrium occupation.
The finite coupling between the system and the bath results in an entangled, non-separable  quantum state of both, system and environment.

All physical observables of the system are encoded in its reduced density matrix, where the environmental degrees of freedom are traced out.
While the time-evolution of the system with environment is in principle unitary, the system dynamics alone proceeds in a non-reversible way due to decoherence, dephasing, and energy relaxation \cite{Feynman1963,Caldeira1983}.
For weak system-bath couplings and for simple structured baths (with short bath correlation time), the dynamics of the system is amendable to analytical approximations \cite{Redfield1957}.
With increasing system-bath couplings and more complex interactions, these approximations break down \cite{Ishizaki2009c} and numerical solutions are required.
For very strong couplings coherence is quickly diminishing and F\"orster theory provides a suitable approximation.

All numerical methods for solving the open quantum system dynamics exactly, require large computational resources and have been limited in their applicability to small systems.
Examples of numerical tools include the Quasi-Adiabatic Path Integral (QUAPI) \cite{Topaler1993} and time-dependent density-matrix renormalization group (t-DMRG) methods \cite{Prior2010}.
Other approaches work with variants of stochastic Schr\"odinger equations \cite{Suess2014} or Monte-Carlo methods \cite{Olsina2014}.
Tanimura and Kubo showed that for an exponentially decaying bath correlation function the system dynamics can be expanded as Hierarchical Equations of Motions (HEOM) \cite{Tanimura1989}.
The HEOM converges fast for systems with energy gaps smaller than the thermal fluctuations $k_B T$. Initially developed for a few level system in contact with many vibrational modes \cite{Tanimura1989,Tanimura1994}, it has been extended to fermionic environments \cite{Jin2008} and applied to a wide range of electronic systems within the HEOM-QUICK program \cite{Ye2016}.
In addition, HEOM has been used to describe the Spin-Boson model \cite{Tsuchimoto2015} and quantum heat transport \cite{Kato2016a}.
The HEOM method provides linear and non-linear spectra of molecular complexes \cite{Chen2009,Chen2010} and retains non-Markovian effects in the population dynamics in models of light harvesting complexes (LHCs) \cite{Ishizaki2008,Ishizaki2009h}.

Light harvesting complexes are large pigment-protein complexes in charge of the energy transfer from the antenna to the reaction center where the chemical reactions take place. Electronic excitation within the pigments is done in quanta or excitons that move within the complex via dipole-dipole interactions and are subject to dissipation due to vibrations of the protein environment. 
In the past years, there has been enormous progress on the experimental manipulation of photochemical reactions and probing of the subsystems of the photosynthetic apparatus of bacteria and higher plants \cite{Nuernberger2015, Dostal2016}. 
This opens up the possibility of understanding natural photosynthesis \cite {Blankenship2014, Reimers2016}, and moreover the controlled design and fabrication of artificial photosynthetic systems \cite{Scholes2011, Romero2017}.
With advanced laser technology short time-scales have become accessible and put more stringent bounds on theoretical models.
Ultrafast spectroscopic techniques on the femtosecond scale are able to follow in real time the flow of the exciton within the complex and have revealed interesting phenomena, including the dissipation and coherence of the excited states \cite{Brixner2005,Engel2007a}.

The interpretation of experimental spectra is done by fitting the measurements to theoretical models.
Due to computational constraints, the models underlying the fit are typically limited with respect to the included physical processes and in their accuracy.
To move beyond simplified approaches and to address the observations of coherences and reorganizational processes in LHCs calls for more accurate and computationally fast solutions of the exciton dynamics.

\section*{\sffamily \Large METHODOLOGY}

All following computations are done with the HEOM method, which captures the system-environment dissipation and decoherence while retaining non-Markovian effects in a non-perturbative manner.
The Distributed Memory Hierarchical Equations of Motions (DM-HEOM) provides a highly optimized implementation of the HEOM method and runs on many-core graphics processing units (GPUs), CPUs in a single workstations, and also in a distributed memory fashion across up to hundreds of compute nodes.
This allows one to overcome the memory and compute barrier imposed by existing single-node HEOM implementations and makes the DM-HEOM method suitable for computing the properties of LHCs for a wide range of system sizes and temperatures.

We use DM-HEOM to compute different physical observables, including
linear and transient absorption spectra, 
static fluorescence spectra, 
circular dichroism,
and time and frequency resolved two-dimensional spectra.

The article is organized as follows:
First we introduce the Frenkel-exciton model and the details of the the HEOM method. 
We describe the theoretical framework used within DM-HEOM for calculating different types of optical spectra.
Next we provide specific examples for the Fenna-Matthews-Olson LHC and compare the DM-HEOM results with more approximative but commonly used approaches, including full and secular Redfield theories \cite{Redfield1957,May2004,Ishizaki2009c}.
We then discuss the distributed memory and computing approach of DM-HEOM and discuss how it speeds up the calculations for larger systems before giving a summary and conclusions.

\section*{\sffamily \Large Frenkel exciton model with light induced dynamics}
\label{sec:Model}

The description of the non-equilibrium dynamics of LHCs starts with the dynamics induced by incident light.
To illustrate the use of DM-HEOM we consider the interaction of a LHC with laser pulses. 
The dynamics of the electronic excitation is modeled with a Frenkel exciton  Hamiltonian \cite{May2004} which in the presence of an external electric field $H_{\rm field}(t)$ reads
\begin{equation}
\label{eq:h-full}
H(t)=H_{\rm g}+H_{\rm ex}+H_{\rm bath}+H_{\rm ex-bath}+H_{\rm field}(t).
\end{equation}
Here, $H_{\rm g}=\varepsilon_0\ket{0}\bra{0}$ represents the ground state Hamiltonian (ground state energy $\varepsilon_0$), $H_{\rm ex}$ denotes the excitation energies and interactions of the pigments, and $ H_{\rm bath}$ models the effect of the surrounding protein environment coupled by $H_{\rm ex-bath}$.  
The excitonic Hamiltonian $H_{\rm ex}^{\rm site}$ for a system of $N_{\rm sites}$ constituents (``sites'') is parametrized as
\begin{equation}
\label{eq:h-ex} 	
H_{0}^{\rm site} =\sum_{m=1}^{N_{\rm sites}}\varepsilon_m^0\ket{m}\bra{m}+\sum_{n\neq m}J_{mn}\ket{m}\bra{n},\quad
H_{\rm ex}^{\rm site} =H_{0}^{\rm site}
+\sum_{m=1}^{N_{\rm sites}}\sum_{v=1}^{V_m}\lambda_{m,v}\ket{m}\bra{m},
\end{equation}
where we introduce the energy $\varepsilon_{m}=\varepsilon_{m}^{0}+\sum_{v=1}^{V_m}\lambda_{m,v}$, which consists of the zero phonon energy  $\varepsilon_{m}^{0}$ shifted by the reorganization energy $\sum_v\lambda_{m,v}$, and the coupling matrix elements $J_{mn}$.
The total vibrational environment consists of $B=\sum_m V_m$ uncorrelated baths, where $V_m$ baths $H_{{\rm bath},m,v}=\sum\limits_{i}\hbar\omega_{m,v,i}(b_{m,v,i}^{\dagger}b_{m,v,i}+\frac{1}{2})$ of harmonic oscillators of frequencies $\omega_{m,v,i}$, with bosonic creation and annihilation operators $b_{m,v,i}$ are assigned to each pigment site $m$. 
The oscillator displacement of each bath mode $(b^{\dagger}_{m,v,i}+b_{m,v,i})$ is coupled to the exciton system by
\begin{equation}
H_{\rm ex-bath}=\sum_m \ket{m}\bra{m} \otimes \sum_v\sum_i \hbar\omega_{m,v,i} d_{mvi}(b^{\dagger}_{m,v,i}+b_{m,v,i}),
\end{equation}
where $d_{mvi}$ denotes the coupling strength related to the spectral density 
\begin{equation}
J_{m,v}(\omega)=\pi\sum_i \hbar^2\omega_{mvi}^2d_{mvi}^2\delta(\omega-\omega_{mvi}).
\end{equation}
The spectral density is connected to the reorganization energy
\begin{equation}
\lambda_{m,v}=\int_0^\infty \frac{J_{m,v}(\omega)}{\pi\omega}\rmd\omega.
\end{equation}
For DM-HEOM, we implement a superposition of (shifted) Drude-Lorentz spectral densities at each site:
\begin{equation}
\label{eq:spectral_density_DL}
J_{m}(\omega)=\sum_{v=1}^{V_m}
\left(
 \frac{\lambda_{m,v}\omega\nu_{m,v}}{{(\omega-\Omega_{m,v})}^2+\nu_{m,v}^2}
+\frac{\lambda_{m,v}\omega\nu_{m,v}}{{(\omega+\Omega_{m,v})}^2+\nu_{m,v}^2}\right),
\end{equation}
with inverse bath correlation time $\nu_{m,v}^{-1}$.
The parameter $\Omega_{m,v}$ shifts the peak position of the spectral density and allows one to vary the pure dephasing and relaxation processes, while maintaining the reorganization energy $\lambda_{m,v}$ \cite{Kreisbeck2012b,Kramer2014}.

\section*{\sffamily \Large Distributed memory hierarchical equations of motion (DM-HEOM)}\label{sec:HEOM}
\subsection*{\sffamily \large HEOM formalism}

The dynamics of an open quantum system is described by the Liouville-von Neumann equation for the full (system and bath) density matrix $\rho_{\rm tot}(t)$,
\begin{equation}
\label{eq:LNE}
\frac{\partial}{\partial t}\rho_{\rm tot}(t)=-\frac{\rmi}{\hbar}[H(t),{\rho}_{\rm tot}(t)].
\end{equation}
The physical observables of the exciton system are computed from the reduced density matrix ${\rho}(t)$ by taking the partial trace of $\rho_{\rm tot}(t)$ with respect to the bath modes 
\begin{equation}
\label{eq:reduced_density_matrix}
{\rho}(t)={\rm Tr}_{\rm bath}\left[{\rho}_{\rm tot}(t)\right]. 
\end{equation}
We solve Eqs.~(\ref{eq:LNE},\ref{eq:reduced_density_matrix}) with the HEOM method introduced by Tanimura and Kubo \cite{Tanimura1989}, following Ref.~\citenum{Tanimura2006}.
We assume a system with $N_{\rm states}$ coupled to $B$ baths with $K$ Matsubara/Pad\'e modes for every bath.
 Multiple states can share one bath, or multiple baths can couple to the same state (for instance a two-exciton state is coupled to two sites, or all the $V$ baths of the parametrized spectral density in Eq.~(\ref{eq:spectral_density_DL}).

HEOM consists of a hierarchy of equations for a set of complex-valued matrices $\sigma_u$ with $N_{\rm states}^2$ entries.
Each $\sigma_u$ is uniquely identified by an integer tuple $\vec{n}_u$ with $W=B K$ entries of positive (including $0$) integers.
The tuples are built up from all possible integer partitions up to depth $D$ defined by $\sum_{i=1}^{W} n_{u,i}\le D$.
This results in a total number of matrices given by the binomial
\begin{equation}
\label{eq:Nmatrices}
N_{\rm matrices}=
\left(\begin{array}{c}
W+D\\
W
\end{array}\right)
\end{equation}
Within a linear memory layout, the $\sigma_u$ matrices are also addressed by the consecutive numbering $u=0,\ldots,N_{\rm matrices}-1$.
Each matrix element of $\sigma_u$ is possibly linked to other matrices $\{\sigma^+, \sigma^-\}$ by ``$+$'' and ``$-$'' vertices.
The ``$+$''-links are established by taking the tuple $\vec{n}$ and adding one of the 
$W$ permutations of the unit tuple $(1,0,0,0,0\ldots,0)$ of length $W$ to $\vec{n}$.
The ``$+$''-connection is only valid if the resulting $\vec{n}^+$ elements satisfy $\sum_{i=1}^{W} n_{u,i}^+\le D$.
Similarly, for the $-$ connection one of the $W$ permutations of $(1,0,0,0,0\ldots,0)$ is subtracted from $\vec{n}$.
The ``$-$''-connection is only valid if all elements in the resulting $\vec{n}^-$ tuple remain $\ge 0$. 
The graph of all ``$\pm$'' connections is computed in a first step to obtain the mapping $u\rightarrow \{u_{\pm}\}$.
The position $l$ of the unit element within the permutation of $(1,0,0,0,0\ldots,0)$ encodes the bath index $b$ of the addressed bath and its Matsubara mode $k$ by the relation
\begin{equation}
l=(b-1)K+k+1, \quad b=1,\ldots,B \quad k=0,\ldots,K-1.
\end{equation}
The inverse relation becomes using the quotient and remainder of integer divisions
\begin{equation}
(b,k)=(((l-1)\;{\rm div}\; K) + 1,(l-1)\;{\rm mod}\; K).
\end{equation}
The hierarchy equation is expressed as
\begin{eqnarray}
\label{eq:HEOM}
\frac{\partial \sigma_u}{\partial t}
&=&-\frac{\rmi}{\hbar}\left[H, \sigma_u\right] \label{eq:HEOMcommutator}\\
&-&\sigma_u
\sum_{b=1}^B
\sum_{k=0}^{K-1}
n_{u,(b,k)} \gamma(b,k)\\
&-&\sum_{b=1}^B\sum_{s=1}^{S(b)}\left[ \frac{2\lambda_b}{\beta\hbar^2\nu_b} - \sum_{k=0}^{K-1} \frac{c(b,k)}{\hbar\gamma(b,k)} \right]{\rm V}_{bs(b)}^\times {\rm V}_{bs(b)}^\times \sigma_u \\
&+&\sum_{b=1}^B
\sum_{s=1}^{S(b)}
\sum_{k=0}^{K-1}
\rmi {\rm V}_{bs(b)}^\times \sigma^+_{(u,b,k)} \\
&+&
\sum_{b=1}^B
\sum_{s=1}^{S(b)}
\sum_{k=0}^{K-1} 
n_{u,(b,k)} \theta_{{\rm MA}(b,k)} \sigma^-_{(u,b,k)}\\
\theta_{\rm MA}(b,k)&=&
\rmi \; c(b,k) {\rm V}_{bs(b)}^\times+\delta_{k,0}\frac{\lambda_b \nu_b}{\hbar} {\rm V}_{bs(b)}^\circ,
\end{eqnarray}
with the definitions
\begin{eqnarray}
S(b)&=& \text{number of states coupled to bath $b$}\\
\gamma(b,k)&=&\left\{
\begin{array}{ll}
\nu_b & k=0\\
&\\
2\pi k/(\beta\hbar)& k>0
\end{array}\right.\\
c(b,k)&=&\left\{
\begin{array}{ll}
\nu_b\lambda_b \cot(\beta\hbar\nu_b/2) & k=0\\
&\\
\frac{4\lambda_b\nu_b}{\beta\hbar}\frac{\gamma(b,k)}{{\gamma(b,k)}^2-\nu_b^2}& k>0
\end{array}\right.\\
bs(b)&&\text{map bath $b$ to state $s$}\\
n_u(b,k)&&\text{gives the $l=(b-1)K+k+1$ entry in tuple $u$},
\end{eqnarray}
and
\begin{eqnarray}
\left({\rm V}_{s}^\times A\right)_{ij}&=&(\delta_{i,s}-\delta_{s,j})A_{ij}\\
\left({\rm V}_{s}^\times {\rm V}_{s}^\times A\right)_{ij}&=& (1-\delta_{i,j}) (\delta_{i,s}+\delta_{s,j})A_{ij}\\
\left({\rm V}_{s}^\circ A\right)_{ij}&=& (\delta_{i,s}+\delta_{s,j})A_{ij}\;.
\end{eqnarray} 
The top hierarchy element 
\begin{equation}
\rho(t)\equiv \sigma_0(t)
\end{equation}
coincides with the reduced density matrix and encodes the exciton dynamics, while the rest of the hierarchy matrices $\sigma_u$ ($u\ge 1$) are called auxiliary density operators (ADOs).

DM-HEOM replaces the Matsubara expansion of the Bose-Einstein distribution 
$\nu_{k}=\frac{2\pi k}{\beta\hbar}$
by a faster converging Pad\'e expansion \cite{Hu2010a} based on the diagonalization of two matrices
\begin{eqnarray}
(\Lambda)_{m,n}&=&\frac{\delta_{m,n-1}}{\sqrt{(2 m+1) (2 n+1)}}
+\frac{\delta_{m,n+1}}{\sqrt{(2 m+1) (2 n+1)}}, \quad m,n=1,\dots 2K\\
(\Lambda')_{m,n}&=&\frac{\delta_{m,n-1}}{\sqrt{(2 m+3) (2 n+3)}}+\frac{\delta_{m,n+1}}{\sqrt{(2 m+3) (2 n+3)}}, \quad m,n=1,\dots 2K-1
\end{eqnarray}
and determining
\begin{equation}
\eta_i=\left(K^2+\frac{3}{2}K\right)
\frac{\prod_{j=1}^{K-1} (\zeta_j^2-\xi_j^2)}
{\prod_{j=1}^K (\xi_i^2-\xi_j^2+\delta_{i,j})}, \quad i=1,\ldots,K
\end{equation}
from the list of decreasing eigenvalues
\begin{eqnarray}
\xi_i&=&\frac{2}{\text{eigenvalue}_i(\Lambda)}, \quad i=1,\ldots K\\
\zeta_i&=&\frac{2}{\text{eigenvalue}_i(\Lambda')}, \quad i=1,\ldots K-1.
\end{eqnarray}
Within the HEOM equation, the switch from the Matsubara to the Pad\'e expansion requires to replace $\gamma(b,k)$ and $c(b,k)$ by:
\begin{eqnarray}
\gamma(b,k)&=&\left\{
\begin{array}{ll}
\nu_b & k=0\\
&\\
\xi_k/(\beta\hbar)& k>0
\end{array}\right.\\
c(b,k)&=&\left\{
\begin{array}{ll}
\frac{2\lambda_b}{\beta\hbar}\left(1-\sum_{j=1}^K\frac{2\eta_k\nu_b^2}{{(\xi_k/(\beta\hbar))}^2-\nu_b^2}\right) & k=0\\
&\\
\frac{4\lambda_b\nu_b}{\beta\hbar}\frac{\eta_k\xi_k/(\beta\hbar)}{{(\xi_k/(\beta\hbar))}^2-\nu_b^2}& k>0
\end{array}\right.\\
\end{eqnarray}
The different hierarchy layers correspond to higher order time derivatives of the ADOs and quickly take numerical large values.
To counter this effect and achieve a more uniform numerical range of all ADOs, we apply the ADO rescaling\cite{Shi2009b} and substitute
\begin{eqnarray}
\sum_{b=1}^B\sum_{s=1}^{S(b)}\sum_{k=0}^{K-1} \rmi {\rm V}_{sb(b)}^\times \sigma^+_{(u,b,k)} &\rightarrow &\sum_{b=1}^B\sum_{s=1}^{S(b)}\sum_{k=0}^{K-1} \rmi \sqrt{(n_{u,(b,k)}+1)|c(b,k)|} {\rm V}_{sb(b)}^\times \sigma^+_{(u,b,k)} \\
\sum_{b=1}^B\sum_{s=1}^{S(b)}\sum_{k=0}^{K-1} n_{u,(b,k)} \theta_{{\rm MA}(b,k)} \sigma^-_{(u,b,k)} & \rightarrow & \sum_{b=1}^B\sum_{s=1}^{S(b)}\sum_{k=0}^{K-1} \sqrt{n_{u,(b,k)}/|c(b,k)|} \theta_{{\rm MA}(b,k)} \sigma^-_{(u,b,k)}.
\end{eqnarray}

\subsection*{\sffamily \large HEOM memory and compute requirements}
\label{subsec:heom_memory}

\begin{figure}[b]
\begin{center}
\includegraphics[draft=false,width=0.6\textwidth]{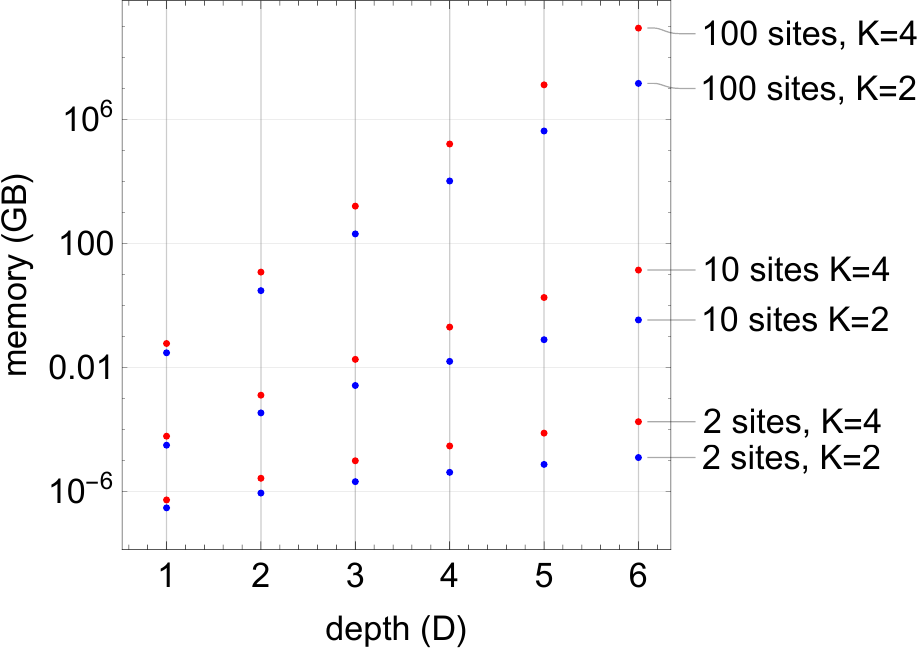}
\end{center}
\caption{Memory requirement for HEOM for increasing truncation depth $D$ 
and increasing system size $N_{\rm states}=\{2,10,100\}$, and different number of Pad\'e modes $K=2,4$.
}\label{fig:heom_memory}
\end{figure}

The HEOM map the exact solution of the open quantum system dynamics to an infinite hierarchy of ADOs.
For practical computations HEOM is evaluated at a finite truncation depth $D$ and for a finite number of Pad\'e modes $K$ at the expense of a (small) numerical error.
The truncation at a finite layer works similar to a Taylor expansion of the time-derivative of the density matrix, where higher order derivatives (corresponding to a deeper layer $D$) contribute with less weight.
The truncated HEOM equations consist of $N_{\rm matrices}=(W+D)!/(W! D!)$ ADOs represented by matrices with $N_{\rm states}^2$ complex-valued floating point numbers which are stored in memory.
For systems with more than 100~states (as found in molecular supercomplexes), the available memory of a single compute node is exhausted (Fig.~\ref{fig:heom_memory}) and it is mandatory to distribute the memory allocation and computation across several nodes.
Low temperature calculations also require of high number of Pad\'e modes and thus a DM-HEOM implementation.

\subsection*{\sffamily \large HEOM convergence and accuracy}
\label{subsec:frobenius_norm}

To study the impact of the truncation level on the accuracy of the results, we analyze the error of the truncated solution with respect to a reference computation at highest feasible truncation level.
This systematic study provides guidelines for choosing the appropriate HEOM depth $D$ and the number of Pad\'e modes $K$ to guarantee a prescribed numerical accuracy of the different spectroscopic quantities. 
The deviations of HEOM from the exact solution can also be studied analytically by analyzing how well HEOM encodes the analytically known line-shape function for a given $(D,K)$ truncation \cite{Kreisbeck2014}.
In addition, systematic error bounds are established in Ref.~\citenum{Mascherpa2017}.
A suitable metric to measure the differences of two matrices is the Frobenius norm of the difference matrix.
The Frobenius norm is defined for a matrix $C$ by
\begin{equation}
\label{eq:frobeniusNorm_deff}
||C||_{\rm F}=\sqrt{{\rm Tr}\{C C^*\}}.
\end{equation}
An exemplary error analysis is carried out in the results section.

\begin{figure}[t]
\includegraphics[draft=false,width=0.9\textwidth]{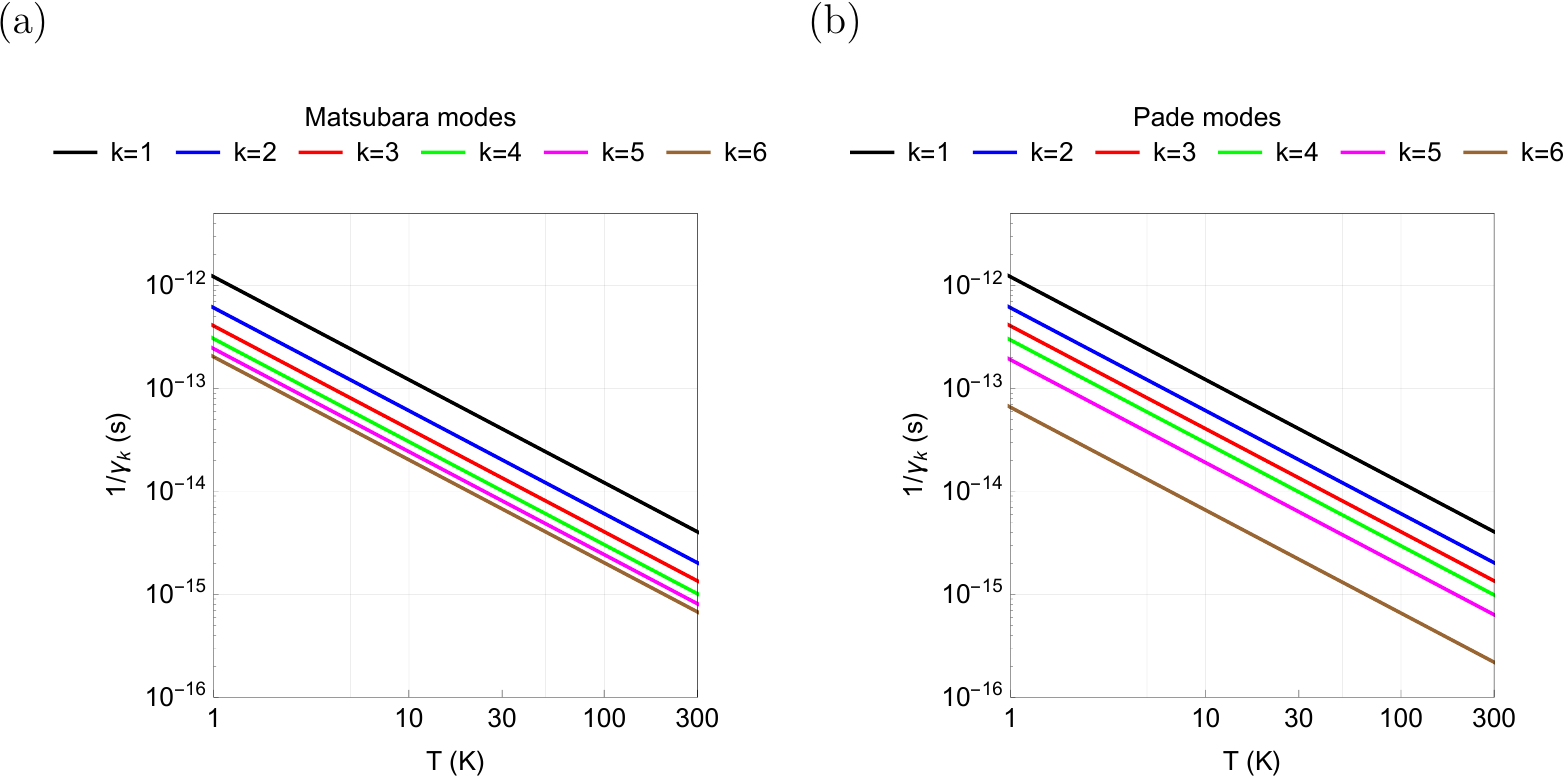}
\caption{
Temperature dependence of the (a) Matsubara and (b) Pad\'e modes.
For a stable numerical result, the integration step-size should be smaller than the period of largest chosen mode at the desired temperature $1/\gamma_k(T)$.
In addition, enough modes must be included to achieve convergence with respect to the exact solution.
}
\label{fig:PadeMatsubara}
\end{figure}

The choice of depth $D$ and Pad\'e modes $K$ determines the largest possible time-step for the Runge-Kutta integration.
Convergence at lower temperatures requires to increase both $D$ and $K$, which results in increased Pad\'e or Matsubara frequencies.
The integration method must resolve these frequencies, which gives an upper limit for the time step $\Delta t$ for each forward time-step:
\begin{equation}
\Delta t\ll 1/\gamma_k.
\end{equation}
Fig.~\ref{fig:PadeMatsubara} shows for every Matsubara and Pad\'e modes the corresponding time periods ($1/\gamma_k$) as function of temperature.

\subsection*{\sffamily \large Redfield approach}
\label{sec:Redfield}

For comparison with commonly used approximations, it is instructive to repeat the computation within the secular and full Redfield approaches.
The full and secular Redfield approaches are given as a closed set of differential equations for the reduced density matrix of a quantum system, but with the known limitation to require a weak system-environment coupling.
The Redfield tensor is usually expressed in the energy representation rather than the site representation used for HEOM.
We denote the unitary transformation between the two basis sets by the diagonalizing matrix $A$
\begin{equation}
H^{\rm exc}={A}\,H_{\rm site}\,{A}^T,
\end{equation}
which leads to a diagonal matrix $H^{\rm exc}$ with $i=1,\ldots,N_{\rm sites}$ eigenenergies $E_i=\hbar\omega_i$.
The Fourier transform of the bath correlation function corresponding to the spectral density in Eq.~(\ref{eq:spectral_density_DL}) is given in terms of the Digammma function $\digamma$:
\begin{eqnarray}
C(\omega)&=&-\frac{\rmi \lambda  \hbar}{2} \bigg[\frac{\nu _+ \left(-\omega +\rmi \nu _-\right) \cot \left(\frac{1}{2} \beta  \nu _+ \hbar \right)+\rmi \nu _- \left(\nu +\rmi
   \omega _+\right) \cot \left(\frac{1}{2} \beta  \nu _- \hbar \right)-2 \left(\nu ^2+\rmi \nu  \omega +\Omega ^2\right)}{\Omega ^2+(\nu +\rmi \omega )^2}\nonumber \\
&& +\frac{\rmi
   \nu _+ \left(\nu ^2+\omega _+^2\right) \digamma \left(\frac{\beta  \nu _+ \hbar }{2 \pi }+1\right)}{\pi  \left(\nu -\rmi \omega _-\right) \left(\nu ^2+\omega _+^2\right)}\\\nonumber
&& +\frac{\rmi \nu _- \left(\nu _+^2+\omega ^2\right) \digamma
   \left(\frac{\beta  \nu _- \hbar }{2 \pi }+1\right)-\rmi \nu _+ \left(\Omega ^2+(\nu -\rmi \omega )^2\right) \digamma \left(1-\frac{\beta  \nu _+ \hbar }{2 \pi
   }\right)}{\pi  \left(\nu -\rmi \omega _-\right) \left(\nu ^2+\omega _+^2\right)}\\\nonumber
&&+\frac{2 \nu  \omega  \left(\frac{1}{\nu ^2+\omega _+^2}+\frac{1}{\nu ^2+\omega
   _-^2}\right) \digamma \left(1+\frac{\rmi \beta  \omega  \hbar }{2 \pi }\right)}{\pi }+\frac{\nu _- \digamma \left(1-\frac{\beta  \nu _- \hbar }{2 \pi
   }\right)}{\pi  \left(-\omega +\rmi \nu _-\right)}\bigg],\\
\nu_\pm&=&\nu\pm\rmi \Omega\\
\omega_\pm&=&\omega\pm\rmi \Omega
\end{eqnarray}
The full Redfield tensor is expressed in terms of the correlation function by
\begin{eqnarray}
R_{\mu\nu\mu'\nu'}&=&\Gamma_{\mu\nu\mu'\nu'}+
{(\Gamma_{\mu\nu\mu'\nu'})}^* -\delta_{\nu\nu'}  \sum_{\kappa=1}^{N_{\rm sites}}\Gamma_{\mu\kappa\kappa\mu'}
-\delta_{\mu\mu'} \sum_{\kappa=1}^{N_{\rm sites}}\Gamma_{\nu\kappa\kappa\nu'},\\
\Gamma_{\mu\nu\mu'\nu'}&=&\frac{1}{\hbar^2}
\sum_{m=1}^{N_{\rm sites}}A_{\mu m}A_{\nu m}A_{\mu' m}A_{\nu' m}
C(\omega_{\nu'}-\omega_{\mu'})
\end{eqnarray}
For the secular Redfield approximation all entries are set to zero which fulfill $(\omega_\mu-\omega_\nu)=(\omega_{\mu'}-\omega_{\nu'})$.
The time evolution of the density matrix elements $\rho_{\mu\nu}$ in energy representation of the exciton Hamiltonian (\ref{eq:h-ex}) is given by
\begin{equation}
\label{eq:Redfield}
\frac{\partial \rho^{\rm exc}_{\mu\nu}(t)}{\partial t}=-\rmi (\omega_{\mu}-\omega_{\nu})\rho^{\rm exc}_{\mu\nu}(t)+\sum_{\mu'=1}^{N_{\rm states}}\sum_{\nu'=1}^{N_{\rm states}}R_{\mu\nu,\mu'\nu'}\rho^{\rm exc}_{\mu'\nu'}(t)\,.
\end{equation}
The first term in eq.~(\ref{eq:Redfield}) describes the coherent evolution governed by the diagonalized Hamiltonian, while the second term leads to decoherence and relaxation governed by the coupling to the baths.
For comparisons with the reduced density matrix given by HEOM, we transform the Redfield tensor back to the site representation
\begin{equation}
\rho_{\rm Redfield}(t)={A}^T\,\rho^{\rm exc}(t)\,{A}.
\end{equation}

\section*{\sffamily \Large Optical spectra}
\label{sec:Spectra}

Here, we discuss the most commonly used spectroscopy for the characterization of the exciton dynamics in LHCs.
To describe the molecular interaction with the electric field, we start from the dipole  operator\cite{Chen2015}
\begin{equation}
H_{\rm field}(t)=-\sum_p {\mathbf e_p} \cdot\hat{\mathbf \mu} E_p(\mathbf{r},t),
\end{equation}
where ${\mathbf e_p}$ is the unit vector in the Cartesian electric field component $E_p(\mathbf{r},t)$ and the dipole matrix operator is given by $\hat{\mathbf \mu}= \hat{\mathbf \mu}^++\hat{\mathbf \mu}^-$,  where
\begin{equation}
\hat\mu^+=\sum_{a=1}^{N_{\rm sites}}  \mathbf{d}_a |a\rangle\langle 0|\,,
\label{eq:mu_plus}
\end{equation}
\begin{equation}
\hat\mu^-=\sum_{a=1}^{N_{\rm sites}}  \mathbf{d}_a|0\rangle\langle a|\,=(\hat\mu^+)^\dagger.
\label{eq:mu_minus}
\end{equation}
In general, $E(\mathbf{r},t)=E^+(\mathbf{r},t)+ E^-(\mathbf{r},t)$, 
such that $E_p^-(\mathbf{r},t)=(E_p^+(\mathbf{r},t))^*$ and 
\begin{equation}
\label{eq:probe_pulse}
E^+(\mathbf{r},t)=\tilde{E}(t-t_{c})\rme^{\rmi(\omega_c t+\mathbf{k}\mathbf{r})}\,
\end{equation}
where $\tilde{E}(t)$ denotes the pulse envelope, centered at $t_{\rm c}$, $\omega_c$ the carrier frequency, and $\varphi=\mathbf{k}\cdot\mathbf{r}$ is the phase of the laser pulse.

Within the rotating-wave approximation (RWA), the complex valued electric field is combined with the respective excitation and de-excitation parts of the dipole operator \cite{Gelin2013,Chen2015}:
\begin{equation}
\label{eq:h-field-rwa}
H_{\rm field}(t)=-\sum_p {\mathbf e_p}\cdot [\hat{\mathbf \mu}^+ E_p^-(\mathbf{r},t) +\hat{\mathbf \mu}^- E_p^+(\mathbf{r},t)]\end{equation}
The optical spectra can be obtained from the evolution of  the time-dependent optical response of the molecular complex, the non-linear polarization $P(t)$ induced by a single (or a combination of) weak probe laser pulse.  The time-dependent polarization is given by
\begin{equation}
P(t)={\rm Tr}[{\rho}(t) \hat{\mu}^+], \quad \rho(t=0)=|0\rangle\langle 0|
\label{eq:polarization}
\end{equation}
where $\rho(t)$ denotes the time-evolved density matrix from the time-dependent Hamiltonian (\ref{eq:h-full}). The trace is taken with respect to the system and bath. 
For weak laser pulses the polarization function can be expanded in powers of the electric field \cite{Hamm2011} and written as a convolution of the electric field with the response function $S^{(n)}(t_n,..,t_1)$ or calculated using a non-perturbative approach. 

\subsection*{\sffamily \large Dipole operators and rotational averaging}

The computation of spectra requires to specify the dipole operator, which accounts for the charge redistribution in the presence of an external electric field in each molecule in the complex ${\bf d}_m$. For short pulses it is a $N_{\rm sites}+1$ dimensional matrix vector Eq.~(\ref{eq:mu_plus}), that reads for each direction $p$,
\begin{equation}
\hat\mu^+_p=\sum_{m=1}^{N_{\rm sites}}  \mathbf{e}_p \cdot \mathbf{d}_m |m\rangle\langle 0|\,.
\label{eq:mu_plusp}
\end{equation}
For longer pulses or multiple short pulses, the excitation of an additional exciton is possible and requires to extend the dipole representation to the two exciton states, which enlarges the Hamiltonian and dipole matrix to $N_{\rm states}$ entries \cite{Cho2005,Hein2012},
\begin{equation}
N_{\rm states}=1+ N_{\rm sites} + \left[N_{\rm sites} (N_{\rm sites} - 1)\right]/2.
\end{equation}

In typical experiments, an ensemble of randomly oriented molecules with respect to the laser direction is probed.
To simplify the theoretical description, we work in the molecular fixed frame and take the rotational average by integrating over different laser directions ${\bf k}_p$.
For linear spectroscopy which probes the first order response function, rotational averaging can be done by considering three representative electric fields \cite{Hein2012} along the Cartesian unit vectors:
\begin{equation}
{\bf e}_{1}=\{1,0,0\},\quad
{\bf e}_{2}=\{0,1,0\},\quad
{\bf e}_{3}=\{0,0,1\}\,.
\label{eq:laser_direction3}
\end{equation}
For two-dimensional spectra, the rotational averaging becomes more involved due to the four dipole interactions involved.
If all laser pulses are equally polarized, $10$ representative electric field directions along the vertices of a dodecahedron suffice \cite{Hein2012}, while for more complex polarization sequences up to $21$ electric field combinations have to be considered \cite{Gelin2017}.

\subsection*{\sffamily \large Linear absorption spectra}
\label{subsec:absorption}

A general approach to the computation of spectra is to evaluate the time evolution of dipole correlation functions (see the review by Gordon \cite{Gordon1968a} for early references) and after time-propagation to take the Fourier transform to switch to the frequency domain.

For a linear absorption spectra with a short initial excitation, the Fourier transform of the polarization correlation function Eq.~(\ref{eq:polarization}) for the sum over polarization directions ${\mathbf e_p}$ becomes
\begin{equation}
\langle{\rm LA}(\omega)\rangle_{\rm rot}={\rm Re}\sum_p\int^{\infty}_{0}\rmd t \exp(\rmi \omega t){\rm Tr}[\hat\mu_p(t)\hat\mu_p(0){\rho}(0)]\,,
\label{eq:lin_abs_mukamel}
\end{equation}
where the dipole operators are calculated in the interaction picture \cite{Hamm2011}.
The trace operates on the system matrix only, since the trace over the environment is already contained in the reduced density matrix.

The evolution of the dipole matrix and the linear absorption Eq.~(\ref{eq:lin_abs_mukamel}) is calculated here using the HEOM Eqs.~(\ref{eq:HEOM}), with all the ADOs initially set to zero and the initial density matrix at $t=0$ is in the ground state $\rho(0)=\sigma_0(0)=|0\rangle\langle 0|$.

At non-zero temperature, decoherence and relaxation towards the thermal state eventually lead to a vanishing correlation function. 
In this case, it is possible to shorten the numerical propagation time to a finite interval and to pad remaining time-intervals with zero to increase the resolution in the frequency domain after the Fourier transform.

\subsection*{\sffamily \large Static Fluorescence Spectra}
\label{subsec:st_fluor}

To compute the static fluorescence (steady-state emission) spectra, we follow \cite{Shuang2001}, to obtain
\begin{equation}
\label{eq:fluorescence}
\langle {\rm FL}(\omega)\rangle_{\rm rot}=
\sum_{p=1}^{3}{\rm Re} \int_0^\infty \rmd t \exp(\rmi \omega t){\rm Tr}
[\hat\mu_p^-(t) \hat{\mu}_p^+ (\infty)\sigma_0^*(\infty)].
\end{equation}
This expression looks similar to the one for the linear absorption, in particular all dipole operations affect all the ADOs.
However, the initial density matrix and ADOs differ from linear absorption, since fluorescence emission starts from the thermal state of the exciton system, 
augmented by the ground state.
The thermal equilibrium state of all ADOs is denoted by $\sigma_u(\infty)$, and can be obtained in two different ways.
Either one propagates the density matrix and ADOs using HEOM for a long time, using as an initial state the Boltzmann distribution function for the reduced density matrix of the system 
\begin{equation}
\label{eq:rho_Boltzmann}
\rho_{\rm Boltzmann}=\rme^{-H_{0}^{\rm site}/k_{\rm B}T} /
{\rm Tr}[ \rme^{-H_{0}^{\rm site}/k_{\rm B}T} ],
\end{equation} 
where $H_{0}^{\rm site}$ denotes the exciton Hamiltonian with the site dependent reorganization energies subtracted Eq.~(\ref{eq:h-ex}).
An alternative method to faster drive HEOM towards the thermal state is the thermal state search method \cite{Zhang2017}.
Both alternatives result in an entangled system-bath state $\sigma_u(\infty)$ differing from the simple Boltzmann distribution state.
This is inherent to the non-separability between the vibrational and electronic modes of the HEOM \cite{Dijkstra2010}. 

The computation of the static fluorescence with the Redfield approach is simplified for the secular Redfield case (which assumes separable system and environment), since there the thermal state takes the Boltzmann value (\ref{eq:rho_Boltzmann}).
For full Redfield, the thermal state needs to be obtained in a separate computation.
The possible violation of positivity by the Full Redfield approach makes it less useful for computing fluorescence spectra, as shown in the results section for the FMO complex.

\subsection*{\sffamily \large Circular dichroism spectra}
\label{subsec:circ_dichr}

The circular dichroism spectrum differs from the linear absorption spectra only by the definition of the excitation dipole matrix, with $\hat{\mu}_p^-$ replaced by
\begin{equation}\label{eq:CDdipoleoperator}
{\hat m}^-_p=\sum_{a=1}^{N_{\rm sites}}  ({\mathbf R}_a\times{\mathbf d}_a)\cdot {\mathbf e}_p |0\rangle\langle a|,
\end{equation}
where the rotational moment is given by the cross product of the radius vector to the center of the $a$th pigment ${\mathbf R}_a$ and the transition dipole moment.
The time evolution of the density matrix in the zero exciton ground state and then excited with the operator eq.~(\ref{eq:CDdipoleoperator}) yields
\begin{equation}\label{eq:LAinit}
\hat{m}^+_p(0)={\hat m}^+_p |0\rangle\langle 0|, \quad \hat{m}^+_p(t)\equiv \sigma_0(t),
\end{equation}
which gives the rotationally averaged circular dichroism
\begin{equation}
\label{eq:cd_spactra_averaged}
\langle {\rm CD}(\omega)\rangle_{\rm rot}=
\sum_{p=1}^{3}{\rm Re} \int_0^\infty \rmd t \exp(\rmi \omega t){\rm Tr}
[\hat\mu_p^- \hat{m}^+_p(t)],
\end{equation}
In the framework of Redfield (secular or full) approach, the averaged circular dichroism spectra is computed as described for the linear absorption case with the same substitution $\hat{\mu}(0) \rightarrow \hat{m}(0)$ for the excitation operator.

\subsection*{\sffamily \large Transient Absorption Spectra}
\label{subsec:TA}

The transient absorption spectra is measured using a  pump-probe laser scheme, where a finite pump pulse $E_{\rm pu}$ prepares a non stationary state, which is monitored by the time-delayed $\tau_{\rm del}$ weak probe pulse $E_{\rm pr}$. 
The TA spectra is obtained from the third order response function using the non-perturbative approach \cite{Seidner1995,Wolfseder1997,Kramer2017a}
\begin{equation}
\label{eq:dif_absorption}
TA (\omega,\tau_{\rm del})=
2\,\omega_{\rm pr}\,{\rm Im}[{\cal E}_{\rm pr}(\omega) 
(\bar{\cal P}^{*}(\omega)-{\cal P}_{\rm only~pr}^{*}(\omega))],
\end{equation}
where we use the Fourier transformed polarization ${\cal P}(\omega)$ and electric field ${\cal E}(\omega)$. 
For a heterodyne phase averaged detection scheme, four propagations of the initial density matrix with different phases of the pump field are required to calculate the non-linear polarization 
\begin{equation}
P(t)=\sum_p{\rm Tr}[\rho(t) \hat{\mu}_p^+].
\end{equation}
The dipole operator includes the two-exciton manifold, which gives rise to excited state absorption (ESA).
The phase of the probe field is set to zero \cite{Seidner1995}, (3.13a):
\begin{equation}
\label{eq:phaseav}
\bar{P}(t)=\frac{1}{4}\big[
P(t,\varphi_{\rm pu}=0)+
P(t,\varphi_{\rm pu}=\frac{\pi}{2})
+P(t,\varphi_{\rm pu}=\pi)+
P(t,\varphi_{\rm pu}=\frac{3\pi}{2})\big].
\end{equation}

\subsection*{\sffamily \large 2D spectra}\label{subsec:2d}

In two-dimensional spectroscopy, a separation of the third order response function along two frequency axes is obtained by taking the Fourier transform along the $t_1$ and $t_3$ time intervals, while the central interval $t_2$ (delay time) is kept parametrically fixed \cite{Mukamel1995,Hamm2011}.
The computation of 2D spectra within the HEOM formalism is described in \cite{Hein2012,Chen2011}.
Here, we consider in addition the possibility of more complicated polarization sequences, which enhance specific processes.

The computation of two-dimensional spectra is demanding due to the need to propagate the density matrix from $t_0=0$ to times $t_1$, $t_2$, and $t_3$.
The corresponding time intervals are $T_1=t_1$, $T_2=t_2-t_1$, $T_3=t_3-t_2$.
In the impulsive limit, the 2D spectra are written in terms of six possible pathways, three rephasing 
\begin{eqnarray}
S_\text{GB,RP} (T_3,T_2,T_1|p_0,p_1,p_2,p_3)&=&+\rmi\,{\rm Tr}\big[{\hat\mu}_{p_3}^-(t_3){\hat\mu}_{p_2}^+(t_2)\rho_0{\hat\mu}_{p_0}^-(0){\hat\mu}_{p_1}^+(t_1)\big]\\
S_\text{SE,RP} (T_3,T_2,T_1|p_0,p_1,p_2,p_3)&=&+\rmi\,{\rm Tr}\big[{\hat\mu}_{p_3}^-(t_3){\hat\mu}_{p_1}^+(t_1)\rho_0{\hat\mu}_{p_0}^-(0){\hat\mu}_{p_2}^+(t_2)\big]\\
S_\text{ESA,RP}(T_3,T_2,T_1|p_0,p_1,p_2,p_3)&=&-\rmi\,{\rm Tr}\big[{\hat\mu}_{p_3}^-(t_3){\hat\mu}_{p_2}^+(t_2){\hat\mu}_{p_1}^+(t_1)\rho_0{\hat\mu}_{p_0}^-(0)\big]
\end{eqnarray}
and three non-rephasing ones
\begin{eqnarray}
S_\text{GB,NR} (T_3,T_2,T_1|p_0,p_1,p_2,p_3)&=&+\rmi\,{\rm Tr}\big[{\hat\mu}_{p_3}^-(t_3){\hat\mu}_{p_2}^+(t_2){\hat\mu}_{p_1}^-(t_1){\hat\mu}_{p_0}^+(0)\rho_0\big]\\
S_\text{SE,NR} (T_3,T_2,T_1|p_0,p_1,p_2,p_3)&=&+\rmi\,{\rm Tr}\big[{\hat\mu}_{p_3}^-(t_3){\hat\mu}_{p_0}^+(0)\rho_0{\hat\mu}_{p_1}^-(t_1){\hat\mu}_{p_2}^+(t_2)\big]\\
S_\text{ESA,NR}(T_3,T_2,T_1|p_0,p_1,p_2,p_3)&=&-\rmi\,{\rm Tr}\big[{\hat\mu}_{p_3}^-(t_3){\hat\mu}_{p_2}^+(t_2){\hat\mu}_{p_0}^+(0)\rho_0{\hat\mu}_{p_1}^-(t_1)\big].
\end{eqnarray}
For a sequence of laser pulses with different relative polarization it is necessary to adjust the electric field directions $p_0$, $p_1$, $p_2$, $p_3$ accordingly.
In addition, an isotropic rotational average of the molecular dipole directions is required for randomly oriented complexes.
We follow Refs.~\citenum{Craig1984,Gelin2017} and implement the tensorial averaging by
\begin{equation}
\langle S(T_3,T_2,T_1) \rangle_{\rm rot}=\sum_{k=1}^{3}\sum_{l=1}^{3}\sum_{m=1}^{3}\sum_{n=1}^{3} C_{klmn} S(T_3,T_2,T_1|p_{0,k},p_{1,l},p_{2,m},p_{3,n}).
\end{equation}
The tensorial average requires to select for the $i$th dipole interaction ($i=0,1,2,3$) a specific Cartesian component $k$ ($k=1,2,3$) of the dipole moment at each pigment:
\begin{eqnarray}
{\hat\mu}_{p_{i,k}}^+&=& 
\sum_{a=1}^{N_{\rm sites}}  \mathbf{e}_{k} \cdot \mathbf{d}_a |a\rangle\langle 0| \\
{\hat\mu}_{p_{i,k}}^-&=&
\sum_{a=1}^{N_{\rm sites}}  \mathbf{e}_{k}  \cdot \mathbf{d}_a |0\rangle\langle a|.
\end{eqnarray}
The factors $C_{klmn}$ are determined by
\begin{eqnarray}
C_{klmn}&=&
\delta_{kl}\delta_{mn}
\left[4(\mathbf{f}_0\cdot\mathbf{f}_1)(\mathbf{f}_2\cdot\mathbf{f}_3)-(\mathbf{f}_0\cdot\mathbf{f}_2)(\mathbf{f}_1\cdot\mathbf{f}_3)-(\mathbf{f}_0\cdot\mathbf{f}_3)(\mathbf{f}_1\cdot\mathbf{f}_2)\right]/30\\\nonumber
&+&
\delta_{km}\delta_{ln}
\left[4(\mathbf{f}_0\cdot\mathbf{f}_2)(\mathbf{f}_1\cdot\mathbf{f}_3)-(\mathbf{f}_0\cdot\mathbf{f}_1)(\mathbf{f}_2\cdot\mathbf{f}_3)-(\mathbf{f}_0\cdot\mathbf{f}_3)(\mathbf{f}_1\cdot\mathbf{f}_2)\right]/30\\\nonumber
&+&
\delta_{kn}\delta_{lm}
\left[4(\mathbf{f}_0\cdot\mathbf{f}_3)(\mathbf{f}_1\cdot\mathbf{f}_2)-(\mathbf{f}_0\cdot\mathbf{f}_1)(\mathbf{f}_2\cdot\mathbf{f}_3)-(\mathbf{f}_0\cdot\mathbf{f}_2)
(\mathbf{f}_1\cdot\mathbf{f}_3)\right]/30,
\end{eqnarray}
where $\mathbf{f}_i$ denotes the unit vector of the electric field of the $i$th pulse $p_i$.
Symmetry reduces the $3^4=81$ $C_{klmn}$ terms to a maximum of $21$ non-zero terms, which are further reduced for specific polarization sequences.

The ESA pathways access the two-exciton manifold \cite{Cho2005,Hein2012}, which enlarges the number of states to propagate from $1+N_{\rm sites}$ to $1+N_{\rm sites}+N_{\rm sites}(N_{\rm sites}-1)/2$
and increase the time required to compute the commutator and the bath interactions considerably.

To obtain the time and frequency resolved 2D spectra for a specific delay time $T_2=(t_2-t_1)$, $S(T_3,T_2,T_1)=S_{\rm RP}+S_{\rm NR}$ is computed separately for the three rephasing (RP) and non-rephasing (NR) pathways for equidistantly spaced times $T_1=0,\Delta t,\ldots,t_1$ and $T_3=0,\Delta t,\ldots,(t_3-t_2)$ and Fourier transformed with different $\omega_1$ signs according to
\begin{eqnarray}
S_{\rm RP}(\omega_3,T_2,\omega_1)    &=&\int_0^\infty\rmd T_1 \int_0^\infty\rmd T_3\,
\rme^{-\rmi T_1\omega_1+\rmi T_3\omega_3} S_{\rm RP}(T_3,T_2,T_1)\\
S_{\rm NR}(\omega_3,T_2,\omega_1)&=&\int_0^\infty\rmd T_1 \int_0^\infty\rmd T_3\,
\rme^{+\rmi T_1\omega_1+\rmi T_3\omega_3} S_{\rm NR}(T_3,T_2,T_1).
\end{eqnarray}
Transient absorption and 2D spectra are related in the impulsive limit via
\begin{equation}
{\rm TA}^{{\rm impulsive}}(\omega,T_2)={\rm Re} \int_{-\infty}^{\infty} d \omega_1 S(\omega,T_2,\omega_1).
\end{equation}
The last relation can be used to validate results from the two approaches.

\section*{\sffamily \Large RESULTS}

\section*{\sffamily \Large Fenna-Matthews Olson complex (FMO)}\label{sec:FMO}

One of the first applications of HEOM to light harvesting complexes has been the study of the exciton population dynamics in the Fenna-Matthews Olson complex \cite{Ishizaki2009h} and its optical properties \cite{Chen2011,Hein2012}.
The FMO is one of the few LHCs where the structural and electronic properties are  well parametrized. A large body of experimental spectra has been published.
This singles out the FMO as one of the simplest LHCs to compare theory and experiments \cite{Reimers2016a}.
DM-HEOM provides a unified framework for computing all optical spectra for a broad temperature range ($30$~K-$300$~K).
In particular the low temperature application of HEOM has been difficult before, since the increasing number of Matsubara terms quickly exhausts the available memory and prolongs the computations. 

In the following, we consider the seven pigment model of the Fenna-Matthews-Olson (FMO) complex parametrized by the following Hamiltonian \cite{Adolphs2006a}
\begin{equation}
 H_{\rm ex}^{\rm site}=
  \left({
  \begin{array}{ccccccc}
   12410 & -87.7 & 5.5  & -5.9 & 6.7 & -13.7 & -9.9 \\
   -87.7 & 12530 & 30.8 & 8.2 & 0.7 & 11.8 & 4.3 \\
    5.5 & 30.8 & 12210 & -53.5 & -2.2 & -9.6 & 6.0 \\
    -5.9 & 8.2 & -53.5 & 12320 & -70.7 & -17.0 & -63.3 \\
     6.7 & 0.7 & -2.2 & -70.7 & 12480 & 81.1 & -1.3 \\ 
     -13.7 & 11.8 & -9.6 & -17.0 & 81.1 & 12630 & 39.7 \\ 
     -9.9 & 4.3 & 6.0 & -63.3 & -1.3 & 39.7 & 12440      
   \end{array} }
   \right)\,\text{cm$^{-1}$.}
\label{eq:Hamiltonian-site-basis-H-lambda}
\end{equation}
The arrangement of the seven pigments and further parameters are listed in Table~\ref{tab:FMOdipbath}.
To facility a comparison of other theories and methods with the HEOM reference calculation, we do not consider static disorder.

\subsection*{\sffamily \large Convergence analysis}\label{subsec:convergence}

We start by establishing the numerical convergence of the DM-HEOM method applied to the FMO complex from a long-time population dynamics up to $t_{\rm max}=10~$ps.
The reference case is provided by the reduced density matrix $\rho(t)^{D=6,K=6}$ calculated with truncation depth $D=6$ and Pad\'e number $K=6$.
For the FMO complex we have computed as example one bath is assigned to each of the pigment sites $B=N_{\rm sites}=7$.
This results in $N_{\rm matrices}=12271512$ matrices in the hierarchy (Eq.~(\ref{eq:Nmatrices})), which are propagated efficiently in parallel.
The deviation of the lower truncation levels to the reference computation is provided by the Frobenius norm Eq.~(\ref{eq:frobeniusNorm_deff}) denoted by
\begin{equation}
||\Delta \rho^{(D,K)}||_{\rm F}=||\rho(t)^{(D,K)}-\rho(t)^{(6,6)}||_{\rm F}.
\label{eq:frobeniusNorm_fin}
\end{equation}

Fig.~\ref{fig:Frobenius_norm_vs_M_T300K_T30K} shows the function $||\Delta \rho^{(D,K)}||_{\rm F}$ for various values of the hierarchy truncation depth $D$ and Pad\'e nodes $K$ at three different temperatures ($T=30$~K, $T=77$~K, $T=300$~K).
Increasing the truncation depth $D$ from $2$ to $6$ in conjunction with increasing $K$ moves the results closer to the reference case. We observe that the requirement in depth truncation  $D$ is more stringent that for the number of Pad\'e nodes $K$, specially at high temperatures. 
In the calculations of optical spectra presented in the following sections, we take as working accuracy results differing less than $10^{-2}$ from the reference computation.
This establishes $D=3$, $K=1$ for $T=300$~K, $D=3$, $K=2$ for $T=77$~K, and $D=3$, $K=4$ for $T=30$~K.

The time step $\Delta t$ used for the integration depends on the Pad\'e mode.
For instance, the reference computation with $K=6$  implies a time step of $\Delta t=0.2$~fs at $T=300$~K (Fig.~\ref{fig:PadeMatsubara}).

\subsection*{\sffamily \large Linear Absorption Spectra}
\label{subsec:linear_absorption_spectra}

Fig.~\ref{fig:LA_redfield} illustrates the linear absorption spectra Eq.~(\ref{eq:lin_abs_mukamel}) at three different temperatures $T=30$, $77$ and $300$~K computed with HEOM, secular Redfield, and full Redfield theories.
The linear absorption spectra is rotationally averaged over three perpendicular laser directions for a linearly polarized laser, but to facilitate a comparison of the theoretical results no inhomogeneous broadening due to disorder is taken into account. 

The HEOM spectra are computed by propagating the density matrix with a step size 
$\Delta t=2$~fs up to $2$~ps.
Depending on the temperature considered, a shorter propagation time can be chosen, since the polarization signal decays faster at higher temperatures.

The agreement between the three methods increases at low temperatures, 
but the HEOM method shows a different temperature-dependent homogeneous broadening, which leads to narrower spectra at $T=300$~K compared to the secular and full Redfield approaches.

Experimentally measured spectra \cite{Vulto1998,Brixner2005} show a similar trend, but in addition require to add an inhomogeneous broadening on the order of $80$~cm$^{-1}$. 
An analysis of the impact of disorder on the linear absorption spectra of FMO calculated with HEOM has been performed in Ref.~\citenum{Hein2012}.

\subsection*{\sffamily \large Fluorescence Spectra}
\label{subsec:fluorescence_spectra}

We calculate the rotationally averaged static fluorescence spectra (Fig.~\ref{fig:FL_redfield_mukamel}) at low $T=30$~K, intermediate $T=77$~K, and room temperature $300$~K. 
The computation starts from the thermal state as initial condition, see Eq.~(\ref{eq:fluorescence}), which increases the computation time for the effort to determine the thermal equilibrium state.
DM-HEOM also implements a faster method to obtain the (entangled) system-bath thermal state following Ref.~\citenum{Zhang2017}.

At low temperature ($T=30$~K), the lowest FMO state is dominantly occupied, resulting in a single pronounced peak.
The line-shape of the HEOM result differs from the one obtained with Redfield theory.
HEOM encodes (up to truncation errors) the exact line-shape function related to the prescribed bath correlation function \cite{Kreisbeck2014}.
An additional difference between HEOM and Redfield theories is the entanglement between bath and exciton modes in the HEOM thermal state and the deviation from the Boltzmann thermal equilibrium, which becomes more relevant at low temperatures.
 
At $T\le 77$~K, the full Redfield method yields negative fluorescence emission, which is unphysical.
This indicates the known lack of positivity of the full Redfield theory \cite{Davis1998a,Cheng2005}. 
For $T=300$~K, secular and full Redfield methods provide very similar results, which differ both from the HEOM approach, especially in the region of high frequencies. 

\subsection*{\sffamily \large Circular dichroism Spectra}
\label{subsec:circular_dichroism_spectra}

Fig.~\ref{fig:CD_redfield_mukamel} shows the circular dichroism spectra averaged over three perpendicular directions, calculated using HEOM, full and secular Redfield approaches for the FMO example system at different temperatures.
The density matrix was propagated to $3$~ps with a time step of $2$~fs. 
We observe better agreement between the HEOM and secular Redfield approaches, specially at $T=300$~K where the disagreement of the full Redfield approach is very pronounced.
The calculated spectra are similar to the experimental CD spectra of FMO at $6K$ in Ref.~\citenum{Vulto1998}.

\subsection*{\sffamily \large Transient absorption spectra}
\label{sebsec:transient_spectra}

In addition to the static spectra presented before, the HEOM time-dependent propagation method is well suited to compute time-resolved spectra.
To demonstrate the fully time-dependent formalism, we consider the pump-probe laser scheme shown in Fig.\ref{fig:laser_pulse_profile}.

Depending on the pulse duration a selective excitation in a specific energy range is achieved, which determines the initial dynamics.
At later times, the system approaches the thermal equilibrium, and is typically probed by a broad bandwidth pulse to reveal the complete redistribution of the deposited energy.
The multiple, finite pulses, require to consider the possibility of excited state absorption in the system and require to carry out the computation in the enlarged state-space, including the two-exciton states, resulting in 29 states in total for FMO.

Transient absorption spectra Eq.~(\ref{eq:dif_absorption}) for FMO are shown in 
Figs~\ref{fig:tr_abs_diff_t_del}, \ref{fig:tr_abs_diff_pu_width} and \ref{fig:tr_abs_diff_t_del_Experiment} for different parameters of delay times and temperatures.
All spectra are laser-phase averaged Eq.~(\ref{eq:phaseav}) and in addition rotationally averaged over $10$ different orientations of the molecular complex with respect to the laser polarization.

The HEOM system was propagated to $3$~ps with a time step of $0.2$~fs to fully resolve the time-dependent laser field which oscillates with the frequencies in the visible spectrum.
Fig.~\ref{fig:tr_abs_diff_t_del} shows the transient spectra as a function of delay time $\tau_{\rm del}$ and temperature.
By increasing the temperature we observe the blurring of the peaks and a shift of the dynamics to lower frequencies.
The impact of different pump-pulse durations $\tau_{\rm pu}$ is demonstrated in Fig.~\ref{fig:tr_abs_diff_pu_width} for a fixed delay time $\tau_{\rm del}=250$~fs. 
By exciting the system with a narrow in time domain pump laser pulse one covers all the frequency domain (see Fig.~\ref{fig:laser_pulse_profile}) as it includes all eigenvalues of $H_{\rm ex}$.

Experimental observations of transient absorption spectra of the FMO complex at $T=10$~K are presented in Ref.~\citenum{Vulto1999}, Fig.~3A for similar laser pulses considered here.
A comparison of the HEOM simulation at $T=77$~K (Fig~\ref{fig:tr_abs_diff_t_del_Experiment}) shows qualitatively similar dynamics: the peaks at longer wavelengths get populated with increasing delay time due to the thermalization.
In addition at higher wavelength the modulation of the positive signal becomes more pronounced.

\subsection*{\sffamily \large 2D spectra}\label{subsec:FMO_2d_spectra}

2D spectra are one of the computationally most demanding applications for DM-HEOM, since they require to evaluate the third order optical response function systematically along three time axes.
The computation can be parallelized across several parameters, for instance different pathways can be computed independently, as well as differing delay times and rotational averages.
The largest computational part is the evaluation of the excited state absorption, which for FMO requires to propagate a $29\times 29$ density matrix.

Fig.~\ref{fig:2d0000} shows the FMO 2D spectra for a series of increasing delay times $T_2$ at $T=100$~K.
Starting at $T_2=100$~fs cross peaks below the diagonal appear, which get more pronounced at longer delay times.
The appearance of the cross peaks is a signature of energy transfer from higher states towards the thermal occupation probabilities \cite{Brixner2005,Engel2007a}.
The exciton energies are correlated with the spatial arrangement of the FMO bacteriochlorophylls to form an energetic funnel from the antenna to the reaction center \cite{Vulto1998,Adolphs2008,Dostal2016,Kramer2017b}.
FMO 2D spectra for different parametrization of the spectral density (including more localized vibrational modes) are discussed in Ref.~\citenum{Kreisbeck2012b}.
Localized vibrational modes affect the 2D spectra, in particular the ground state bleaching pathway \cite{Kreisbeck2013}, while the pure dephasing time, associated with the slope of the spectral density $J(\omega)$ at $\omega=0$, determines the life-time of electronic coherences \cite{Kreisbeck2012b,Kreisbeck2013}.

The relative contributions of stimulated emission, ground state bleaching, and excited state absorption, as well as the relation between 2D spectra and transient absorption spectra is analyzed in Ref.~\citenum{Kramer2017b} for an enlarged model of the FMO complex.
The energy transfer and relaxation towards lower lying states is directly reflected in the stimulated emission signal.
In addition the stimulated emission signal is off-diagonally shifted to lower emission frequencies $\omega_3$ after the reorganization process takes place.

To demonstrate the impact of pulse sequences with varying polarizations, we consider two laser setups: one with all pulses having the same polarization direction 
$S_V=\{0,0,0,0\}$, and one where the electric field of the first two pulses is rotated $\pi/2$ counter-clockwise around the propagation direction $S_H=\{\frac{\pi}{2},\frac{\pi}{2},0,0\}$.
The 21 $C_{klmn}$ coefficients for this polarization sequence are listed in Table~\ref{tab:Cklmn}.
Fig.~\ref{fig:2dpol} shows the resulting rephasing spectra of the FMO complex at delay time $t_2=40$~fs. 
The $S_H$ polarization sequence enhances the cross-peaks and by choosing the $S_Y$ linear combination of the signals, the diagonal peaks are effectively removed.
Corresponding experimental results for (c,d) by Thyrhaug et al are shown in Fig.~2, Ref.~\citenum{Thyrhaug2016a}.

\section*{\sffamily \Large Distributed memory HEOM implementation}\label{sec:DAS}

Previous implementations of the HEOM equations used many-core processors for the parallel computation of the hierarchy equations using threads on CPUs (PHI-HEOM\cite{Strumpfer2012a}) or on graphics processing units (GPU-HEOM \cite{Kreisbeck2011}).
Apart from efficiency one key goal in the development of the DM-HEOM framework was to provide code portability over several computer architectures ranging from notebooks to GPUs, many-core systems, and supercomputers.
Using the Open Computing Language (OpenCL) allows to share a similar code base for both, CPUs and GPUs (QMaster \cite{Kreisbeck2014}) and thus facilitates the incremental optimization process. 
One important difference between GPUs and CPUs is the overhead to launch a compute thread: CPUs typically perform better with fewer threads (one per core/hardware thread), but more computationally intense ones compared to GPUs which excel at thousands of lightweight threads.

An important figure of merit is the arithmetic intensity, i.e.\ the number of floating point operations (FLOP) of the algorithm compared to the amount of memory (bytes) accessed to perform this computation \cite{Williams2009}.
The commutator term in HEOM (Eq.~\ref{eq:HEOMcommutator}) requires $16 N_{\rm states}^3$ FLOP for each ADO.
To copy the complex valued ADO into memory and writing to it entails $2\times 2\times 8 \times N_{\rm states}^2$~bytes, resulting in an arithmetic intensity of $N_{\rm states}/2$~FLOP/byte.
This number has to be compared to the typical CPU and GPU peak floating point performance divided by the memory bandwidth.
For high end GPUs this value ranges from $2-7$~FLOP/byte, while many-core CPUs reach $7-10$~FLOP/byte.
Both compute architectures are in principle well suited for the HEOM method.

The main limitation of the existing HEOM implementations is the memory limit imposed by the single-node GPU memory or CPU accessible RAM.
Molecular systems with more than 100 sites are exceeding the 100~GB memory threshold (see Fig.~\ref{fig:heom_memory}) of workstations.
The required memory increases rapidly upon inclusion of more Matsubara or Pad\'e modes $K$ or truncation depth $D$.

To move beyond this barrier requires to distribute the data across multiple compute nodes, which are interconnected to exchange results required for the next propagation step.
The ADOs of the different layers of the HEOM equations are represented as vertices in a graph, where the edges encode the links between the ADOs.
DM-HEOM splits the ADOs into self-contained parts and halo regions that are shared between interconnected nodes.
While the compute time decreases ideally in proportion with the number of compute nodes thrown at the problem, the communication time does not decrease beyond a problem-specific number of nodes.
This is due to the large amount of transferred data and the high connectivity between the partitions of the problem.
Even when trying to overlap communication with computation as much as possible, the synchronization overhead eventually limits the scalability of the code in terms of total runtime.
However, it can still be useful to run DM-HEOM with more nodes if the memory requirements of the physical system would otherwise be prohibitive to obtain a result with less nodes.

\begin{figure}[t]
\includegraphics[draft=false,width=0.9\textwidth]{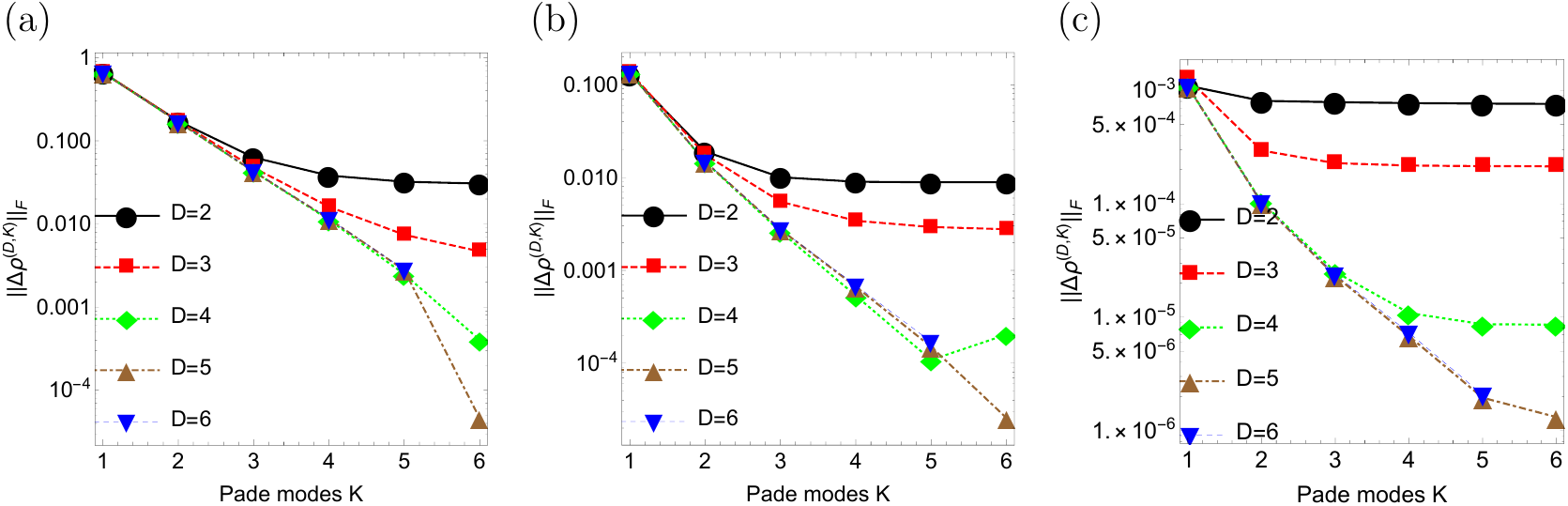}
\caption{
Accuracy of DM-HEOM for increasing truncation level of HEOM.
The error is measured by the Frobenius norm of the difference of the density matrix at truncation levels $(D,K)$ $\rho^{D,K}(t=10$~ps$)$ with respect to the higher order reference $\rho^{6,6}(t=10$~ps$)$.
Temperatures: (a) $T=30~$K, (b) $T=77~$K, (c) $T=300$~K.
}\label{fig:Frobenius_norm_vs_M_T300K_T30K}
\end{figure}

\begin{figure}[t]
\includegraphics[draft=false,width=0.9\textwidth]{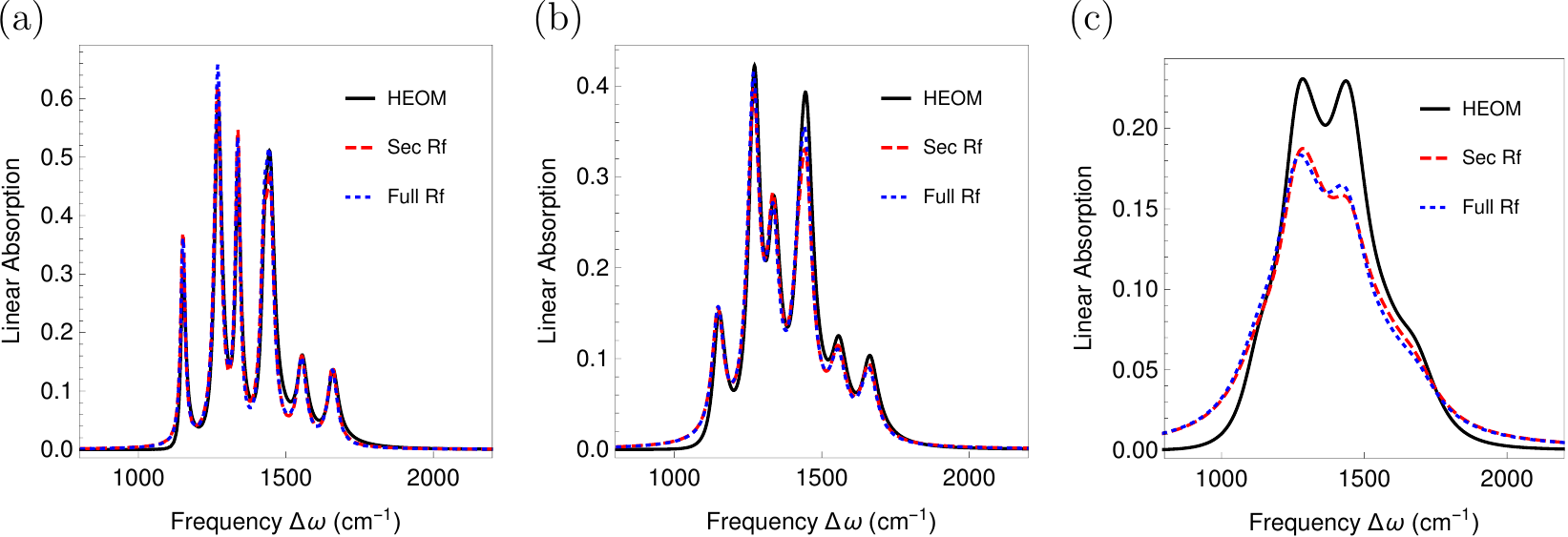}
\caption{
Linear absorption spectra of FMO.
HEOM (black solid), secular Redfield (red dashes), and full Redfield (blue short dashes) theories. 
(a) $T=30$~K (HEOM truncation $D=3$, $K=4$); 
(b) $T=77$~K (HEOM truncation $D=3$, $K=2$); 
(c)  $T=300$~K (HEOM truncation $D=3$, $K=1$). 
}
\label{fig:LA_redfield}
\end{figure}

\begin{figure}[t]
\includegraphics[draft=false,width=0.9\textwidth]{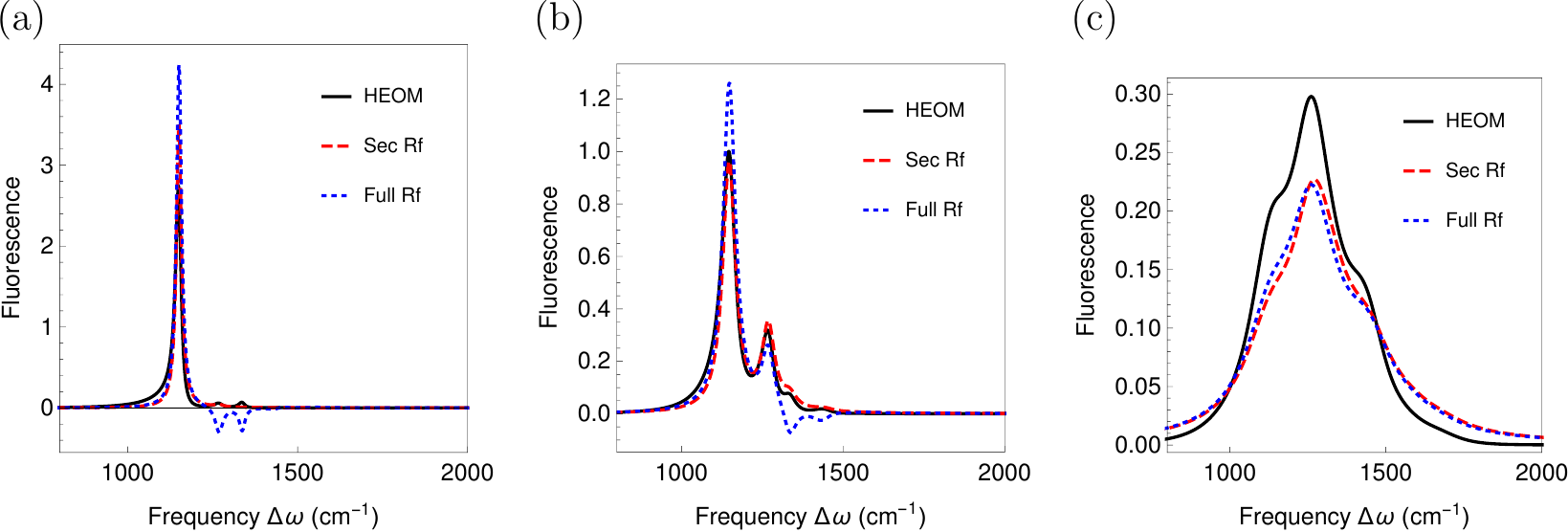}
\caption{
Stationary fluorescence of FMO for temperatures (a) $T=30~$K, (b) $T=77~$K, (c) $T=300$~K.
HEOM (black solid), 
secular Redfield (red dashes), 
full Redfield (blue short dashes) theories. 
Full Redfield theory yields unphysical negative populations at $T=30$~K and $T=77$~K and should not be used to compute fluorescence spectra for these parameters.
}
\label{fig:FL_redfield_mukamel}
\end{figure}

\begin{figure}[t]
\includegraphics[draft=false,width=0.9\textwidth]{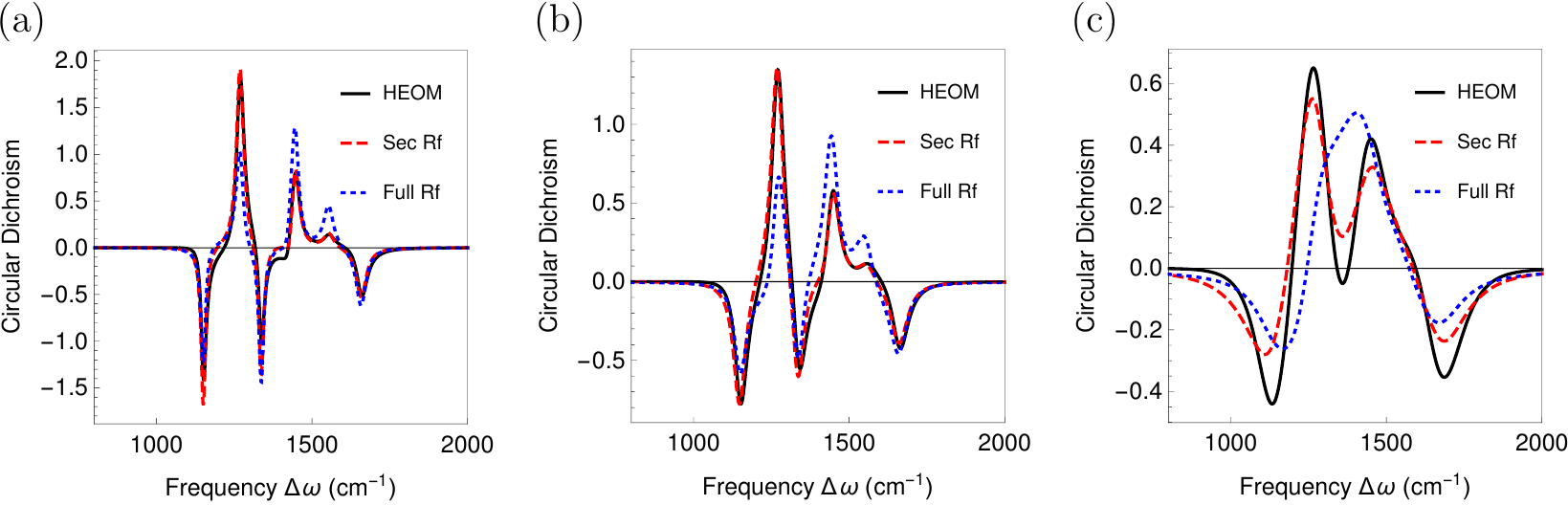}
\caption{Circular dichroism spectra of FMO normalized to its positive area for temperatures (a) $T=30~$K, (b) $T=77~$K, (c) $T=300$~K.
HEOM  (black solid), secular Redfield (red dashes), and full Redfield (blue short dashes) theories are shown.}
\label{fig:CD_redfield_mukamel}
\end{figure}

\begin{figure}[t]
\includegraphics[draft=false,width=\textwidth]{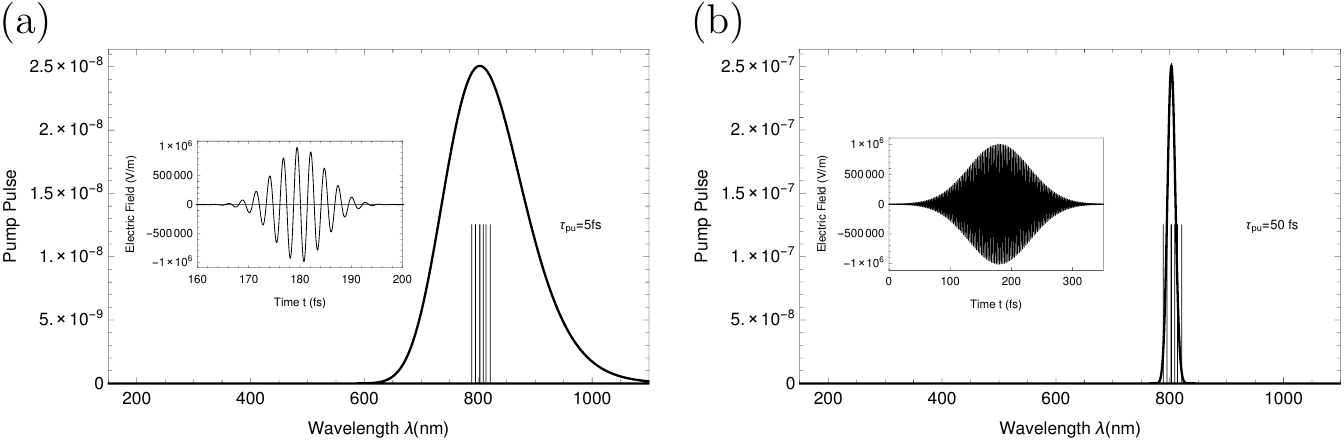}
\caption{Wavelengths covered by the laser pulse 
${E}_{\rm p}(t-t_{\rm c,p})=E_{\rm p}\exp[-(t-t_{\rm c,p})^2/2\tau_{\rm p}]$ for different pulse widths (a) $\tau_{\rm pu}=5$~fs (Full width at half maximum (FWHM) is $11.77$~fs) and (b) $\tau_{\rm pu}=50$~fs FWHM$=117.74$~fs).
The time-dependent electric field amplitude is shown in the inset. 
The pump-probe excitation is applied with 
$E_{\rm pr}=10^{6}$~V/m, 
$E_{\rm pr}=5\times 10^{4}$~V/m, 
$\tau_{\rm pr}=5$~fs and 
$\omega_{\rm pu}=\omega_{\rm pr}=12454.8$~cm$^{-1}$ 
(corresponding to a wavelength of $802.9$~nm). 
The eigenvalues of the FMO Hamiltonian $H_{\rm ex}$ are indicated by the vertical lines.
}
\label{fig:laser_pulse_profile}
\end{figure}

\begin{figure}[t]
\includegraphics[draft=false,width=\textwidth]{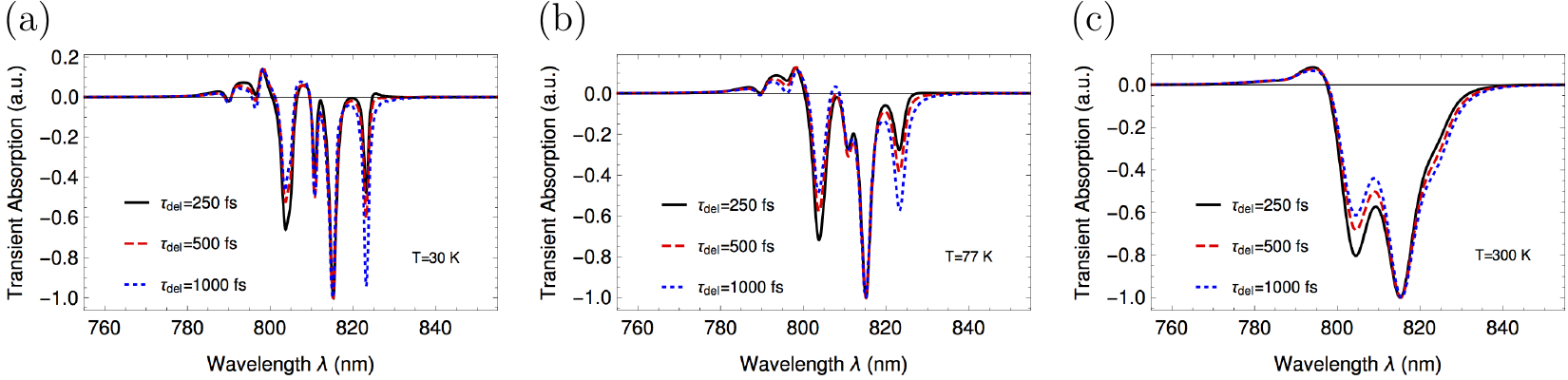}
\caption{Transient absorption spectra of FMO at different delay times and temperatures for $\tau_{\rm pr}=5$~fs (HEOM only) for temperatures (a) $T=30~$K, (b) $T=77~$K, (c) $T=300$~K
}\label{fig:tr_abs_diff_t_del}
\end{figure}

\begin{figure}[t]
\includegraphics[draft=false,width=\textwidth]{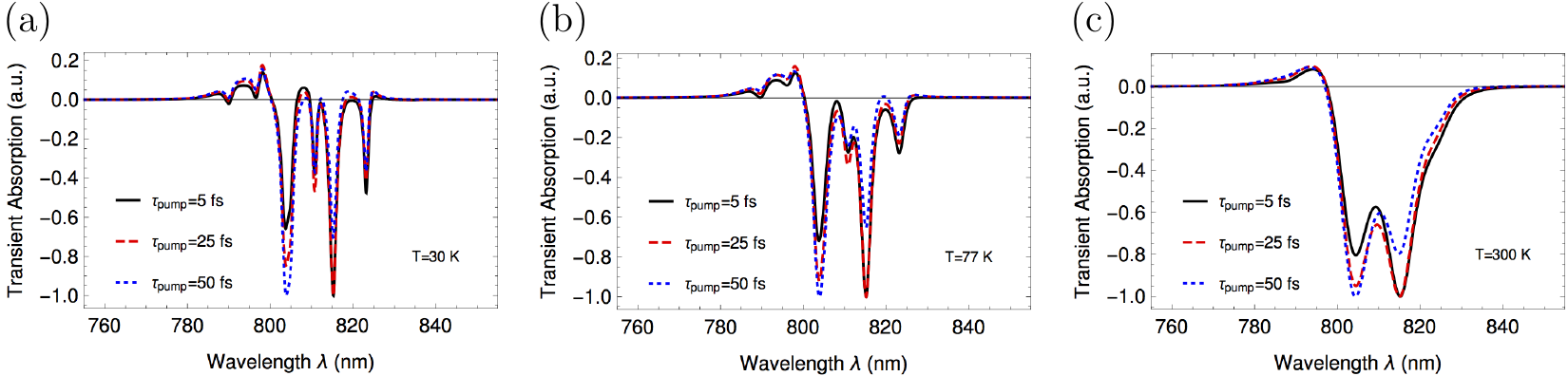}
\caption{Transient absorption spectra of FMO (HEOM only) at different pump laser widths $\tau_{\rm pu}$ and temperatures (a) $T=30~$K, (b) $T=77~$K, (c) $T=300$~K at fixed delay time $\tau_{\rm del}=250$~fs and $\tau_{\rm pr}=5$~fs.}
\label{fig:tr_abs_diff_pu_width}
\end{figure}

\begin{figure}[t]
\begin{center}
\includegraphics[draft=false,width=0.5\textwidth]{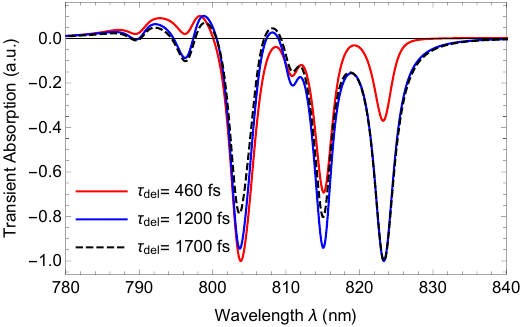}
\end{center}
\caption{
Transient absorption spectra of FMO at different delay times $\tau_{\rm del}$ and $T=77$~K. Computational parameters of the laser pulses are included in caption of Fig.~\ref{fig:laser_pulse_profile}. 
The pump pulse has a width of $\tau_{\rm pu}=128$~fs (corresponding FWHM is $300$~fs as in the experiment \cite{Vulto1999}), the probe pulse width is $\tau_{\rm pr}=5$~fs.
With increasing delay time the peaks at longer wavelengths get more populated.
}
\label{fig:tr_abs_diff_t_del_Experiment}
\end{figure}

\begin{figure}[t]
\includegraphics[draft=false,width=0.9\textwidth]{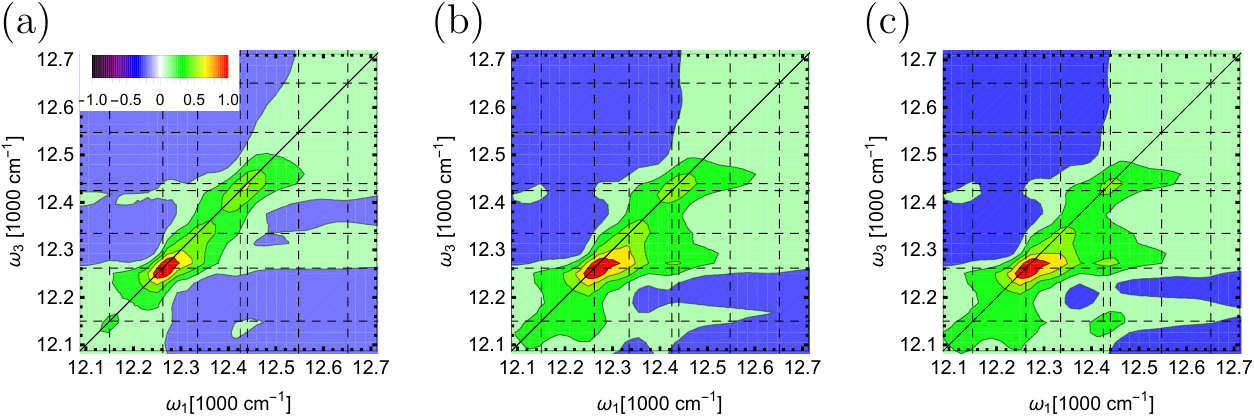}
\caption{
Rotationally averaged FMO 2D spectra (real part of the rephasing signal) at $T=100$~K for increasing delay time (40,100,500)~fs
(HEOM truncation $D=3$, $K=1$).
The color bar is inset in panel (a) (arbitrary units).
}
\label{fig:2d0000}
\end{figure}

\begin{figure}[t]
\includegraphics[draft=false,width=\textwidth]{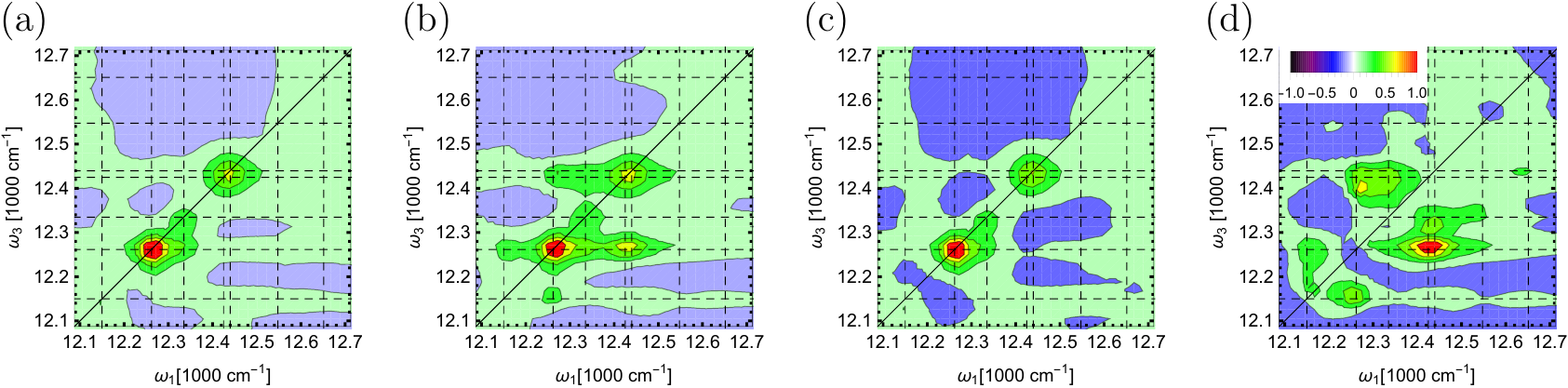}
\caption{
Rotationally averaged FMO 2D spectra (real part of the rephasing and non-rephasing signal) at $T=77$~K for delay time $40$~fs for different polarization sequences. 
(a) $S_V=\{0,0,0,0\}$,
(b) $S_H=\{\frac{\pi}{2},\frac{\pi}{2},0,0\}$,
(c) synthetic: $S_Z=\frac{1}{3}(S_V+2S_H)(5\frac{S_V-S_H}{S_V+2S_H}+1)$,
(d) synthetic: $S_Y=\frac{1}{3}(S_V+2S_H)(2-5\frac{S_V-S_H}{S_V+2S_H})$.
The color bar is inset in panel (d) (arbitrary units).
}
\label{fig:2dpol}
\end{figure}

\begin{figure}[t]
\begin{center}
\includegraphics[draft=false,width=0.5\textwidth]{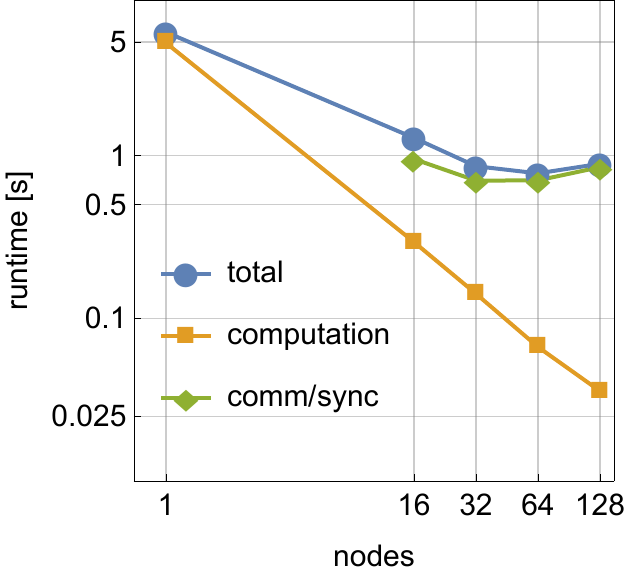}
\end{center}
\caption{Runtime per propagation step for the FMO calculation using $44964388$ ADOs ($K=5$, $D=6$) evaluated on a Cray XC40 supercomputer with Intel Xeon Haswell CPUs. 
By distributing the computation across $16, \ldots, 64$ nodes, the runtime is reduced from $5.8$~s to $0.8$~s on 64 nodes for each propagation step (consisting of 4 HEOM evaluations for the RK4 integration).
The computational part decreases ideally with the number of nodes at the cost of increased communication and synchronization overhead.
As required in such strong scaling measurements, the single node result does not involve any communication overhead.}\label{fig:FMOruntimes}
\end{figure}

To illustrate the reduction in compute time with a distributed run, Fig.~\ref{fig:FMOruntimes} charts the runtime for the FMO population dynamics in Fig.~\ref{fig:Frobenius_norm_vs_M_T300K_T30K} with $D=6$, $K=5$ with
$N_{\rm matrices}=44964388$.
On a single node this computation is feasible, but requires $4\times 3.3=13.1$~GiB memory to store the 4 copies of the ADOs required for an RK4 integration step.
To advance the HEOM system for one $0.2$~fs step (with 4 intermediate results) takes $5.8$~s on a 24~core Intel Xeon Haswell CPU (E5-2680 v3) operated at $2.50$~GHz.
The $50000$ steps propagation to obtain the 10~ps result (Fig.~\ref{fig:Frobenius_norm_vs_M_T300K_T30K}) take $80$~h.
Using DM-HEOM on $16$, $32$ or $64$ nodes, the runtime is reduced to $17.9$~h, $12.0$~h, $10.8$~h respectively.
Increasing the number of nodes to $128$ increases the runtime due to the larger communication overhead.

A fast network connection, as realized on current supercomputers, is essential to maintain best-possible performance for the largest problem sizes considered.
The node distributed results shown here are obtained on the HLRN supercomputing facilities hosting a Cray XC40 with an Aries interconnected network.
The DM-HEOM tools is written such that it runs also on single compute nodes equipped with one or multiple GPUs/CPUs, or across networked compute nodes \cite{Noack2018}.

Other aspects of the evaluation of optical spectra are computed in parallel without additional overhead: 
the rotational averaging and the polarization sequences can be computed independently, cutting down computational times by factors of $3$ or up to $21$ for linear absorption and 2D spectra respectively.

For the accuracy discussed before (relative error $<0.01$) the computation of the FMO dynamics takes about $5.5$~s on a single Intel Xeon Haswell CPU node with $D=3$, $K=1$ up to $10$~ps with 10000 steps. 
The computation of the first order response function for the spectra takes a similar time.
DM-HEOM \cite{Noack2018} is written in C++ and can be easily extended to higher-order optical sequences or other applications of the HEOM equations.
An open source release of DM-HEOM is in preparation, a ready-to-run GPU accelerated HEOM implementation is available at nanoHub.org \cite{GPUHEOM}.

\section*{\sffamily \Large CONCLUSIONS}\label{sec:Discussion}

HEOM is a unique exact method to compute the dynamics in open quantum systems and is frequently used as a benchmark and reference method for more approximative methods, but has also seen limited application to larger systems due to its computational demands. 

In this paper, we have provided a comprehensive review of the HEOM formalism and how it is used in DM-HEOM to efficiently calculate the optical response properties of LHCs. 
We have shown that the DM-HEOM framework provides an accurate and fast implementation of HEOM to compare theoretical models with the most common experimental spectral signals used to characterize light harvesting systems for a new range of parameters.
DM-HEOM extends the applicability of HEOM to lower temperatures ($T=30$~K) and to bigger systems than previously accessible with HEOM. 
We conducted a systematic analysis of the accuracy and convergence of HEOM with respect to the truncation depth and the Pad\'e modes.
This provides a guideline for choosing the optimal time-steps and truncation levels in practical applications, as demonstrated here for the exemplary FMO complex.
The DM-HEOM framework allows one to compute the different optical spectra (linear absorption, fluorescence, and circular dichroism spectra) and to compare them to approximative approaches (here: secular and full Redfield theories) or other exact methods.
Moreover, DM-HEOM implements the efficient calculation of 2D spectra for different polarization sequences and the polarization for finite laser pulses.

The implementation of DM-HEOM overcomes the excessive memory requirements of the HEOM method required for numerical simulations in the very low-temperature regime, which hinders the use of HEOM for investigating quantum phase transitions \cite{Ye2016}. 
Future extensions of DM-HEOM will focus on implementing the Spin Boson variant of HEOM \cite{Tsuchimoto2015}, which allows DM-HEOM to perform calculations at sufficiently low temperatures to model quantum-technology applications. 

\subsection*{\sffamily \large ACKNOWLEDGMENTS}

The work was supported by the German Research Foundation (DFG) grants KR~2889 and RE~1389 (``Realistic Simulations of Photoactive Systems on HPC Clusters with Many-Core Processors'') and the Intel Research Center for Many-core High-Performance Computing at ZIB.
We acknowledge compute time allocation by the North-German Supercomputing Alliance (HLRN).
M.R.\ has received funding from the European Union's Horizon 2020 research and innovation programme under the Marie Sklodowska-Curie grant agreement No.~707636. 
We thank C.~Kreisbeck, M.~Gelin, F.~Mueller, and Th.~Steinke for helpful discussions and J.~Launer, L.~Deecke, and L.~Gaedke-Merzh\"auser for contributing to DM-HEOM.

\clearpage

\begin{table}[b]
\begin{center}
\begin{tabular}{c|c|c|c}
pigment & center (nm) & direction & $\lambda$ cm$^{-1}$, $\nu^{-1}$ (fs), $\Omega$\\\hline
1 & $2.651, +0.260, -1.135$ & $-0.741, -0.561, -0.3696$ & $35$,$50$,$0$\\
2 & $1.560, -0.152, -1.725$ & $-0.857, +0.504, -0.107$ & $35$,$50$,$0$ \\
3 & $0.339, -1.361, -1.385$ & $-0.197, +0.957, -0.211$ & $35$,$50$,$0$ \\
4 & $0.668, -2.085, -0.604$ & $-0.799, -0.534, -0.277$ & $35$,$50$,$0$ \\
5 & $1.938, -1.857, -0.108$ & $-0.737, +0.656, +0.164$ & $35$,$50$,$0$ \\
6 & $2.184, -0.718, +0.063$ & $-0.135, -0.879, +0.457$ & $35$,$50$,$0$ \\
7 & $1.027, -0.821, -0.554$ & $-0.495, -0.708, -0.503$ & $35$,$50$,$0$
\end{tabular}
\end{center}
\caption{Centers and orientations of the FMO dipoles, taken from the PDB:3ENI structure.
The parameters for the spectral density (reorganization energy $\lambda$) and bath correlation time $\nu^{-1}$ are from Ref.~\citenum{Hein2012}.
}\label{tab:FMOdipbath}
\end{table}

\begin{table}[b]
\begin{center}
\begin{tabular}{c|c}
$(k,l,m,n)$ & $C_{klmn}$ \\\hline
$(1, 1, 1, 1)$, $(1, 1, 2, 2)$, $(1, 1, 3, 3)$ & $\frac{1}{15}$,$\frac{2}{15}$,$\frac{2}{15}$ \\
$(1, 2, 1, 2)$, $(1, 2, 2, 1)$, $(1, 3, 1, 3)$ & $-\frac{1}{30}$,$-\frac{1}{30}$,$-\frac{1}{30}$ \\
$(1, 3, 3, 1)$, $(2, 1, 1, 2)$, $(2, 1, 2, 1)$ & $-\frac{1}{30}$,$-\frac{1}{30}$,$-\frac{1}{30}$ \\
$(2, 2, 1, 1)$, $(2, 2, 2, 2)$, $(2, 2, 3, 3)$ & $\frac{2}{15}$,$\frac{1}{15}$,$\frac{2}{15}$ \\
$(2, 3, 2, 3)$, $(2, 3, 3, 2)$, $(3, 1, 1, 3)$ & $-\frac{1}{30}$,$-\frac{1}{30}$,$-\frac{1}{30}$ \\
$(3, 1, 3, 1)$, $(3, 2, 2, 3)$, $(3, 2, 3, 2)$ & $-\frac{1}{30}$,$-\frac{1}{30}$,$-\frac{1}{30}$ \\
$(3, 3, 1, 1)$, $(3, 3, 2, 2)$, $(3, 3, 3, 3)$ & $\frac{2}{15}$,$\frac{2}{15}$,$\frac{1}{15}$ 
\end{tabular}
\end{center}
\caption{$C_{klmn}$ coefficients for isotropic averaging of the $S_H=\{\frac{\pi}{2},\frac{\pi}{2},0,0\}$ polarization sequence.
}\label{tab:Cklmn}
\end{table}

\begin{thebibliography}{65}
\expandafter\ifx\csname natexlab\endcsname\relax\def\natexlab#1{#1}\fi
\expandafter\ifx\csname bibnamefont\endcsname\relax
  \def\bibnamefont#1{#1}\fi
\expandafter\ifx\csname bibfnamefont\endcsname\relax
  \def\bibfnamefont#1{#1}\fi
\expandafter\ifx\csname citenamefont\endcsname\relax
  \def\citenamefont#1{#1}\fi
\expandafter\ifx\csname url\endcsname\relax
  \def\url#1{\texttt{#1}}\fi
\expandafter\ifx\csname urlprefix\endcsname\relax\def\urlprefix{URL }\fi
\providecommand{\bibinfo}[2]{#2}
\providecommand{\eprint}[2][]{\url{#2}}

\bibitem[{\citenamefont{Feynman and Vernon}(1963)}]{Feynman1963}
\bibinfo{author}{\bibfnamefont{R.}~\bibnamefont{Feynman}} \bibnamefont{and}
  \bibinfo{author}{\bibfnamefont{F.}~\bibnamefont{Vernon}},
  \bibinfo{journal}{Annals of Physics} \textbf{\bibinfo{volume}{24}},
  \bibinfo{pages}{118} (\bibinfo{year}{1963}), ISSN \bibinfo{issn}{00034916},
  \urlprefix\url{http://linkinghub.elsevier.com/retrieve/pii/000349166390068X}.

\bibitem[{\citenamefont{Caldeira and Leggett}(1983)}]{Caldeira1983}
\bibinfo{author}{\bibfnamefont{A.}~\bibnamefont{Caldeira}} \bibnamefont{and}
  \bibinfo{author}{\bibfnamefont{A.}~\bibnamefont{Leggett}},
  \bibinfo{journal}{Physica A} \textbf{\bibinfo{volume}{121}},
  \bibinfo{pages}{587} (\bibinfo{year}{1983}), ISSN \bibinfo{issn}{03784371}.

\bibitem[{\citenamefont{Redfield}(1957)}]{Redfield1957}
\bibinfo{author}{\bibfnamefont{A.~G.} \bibnamefont{Redfield}},
  \bibinfo{journal}{IBM Journal of Research and Development}
  \textbf{\bibinfo{volume}{1}}, \bibinfo{pages}{19} (\bibinfo{year}{1957}),
  ISSN \bibinfo{issn}{0018-8646},
  \urlprefix\url{http://ieeexplore.ieee.org/lpdocs/epic03/wrapper.htm?arnumber=5392713}.

\bibitem[{\citenamefont{Ishizaki and
  Fleming}(2009{\natexlab{a}})}]{Ishizaki2009c}
\bibinfo{author}{\bibfnamefont{A.}~\bibnamefont{Ishizaki}} \bibnamefont{and}
  \bibinfo{author}{\bibfnamefont{G.~R.} \bibnamefont{Fleming}},
  \bibinfo{journal}{The Journal of Chemical Physics}
  \textbf{\bibinfo{volume}{130}}, \bibinfo{pages}{234110}
  (\bibinfo{year}{2009}{\natexlab{a}}), ISSN \bibinfo{issn}{00219606},
  \urlprefix\url{http://www.ncbi.nlm.nih.gov/pubmed/19548714
  http://scitation.aip.org/content/aip/journal/jcp/130/23/10.1063/1.3155214}.

\bibitem[{\citenamefont{Topaler and Makri}(1993)}]{Topaler1993}
\bibinfo{author}{\bibfnamefont{M.}~\bibnamefont{Topaler}} \bibnamefont{and}
  \bibinfo{author}{\bibfnamefont{N.}~\bibnamefont{Makri}},
  \bibinfo{journal}{Chemical Physics Letters} \textbf{\bibinfo{volume}{210}},
  \bibinfo{pages}{285} (\bibinfo{year}{1993}), ISSN \bibinfo{issn}{00092614},
  \urlprefix\url{http://linkinghub.elsevier.com/retrieve/pii/0009261493891355}.

\bibitem[{\citenamefont{Prior et~al.}(2010)\citenamefont{Prior, Chin, Huelga,
  and Plenio}}]{Prior2010}
\bibinfo{author}{\bibfnamefont{J.}~\bibnamefont{Prior}},
  \bibinfo{author}{\bibfnamefont{A.~W.} \bibnamefont{Chin}},
  \bibinfo{author}{\bibfnamefont{S.~F.} \bibnamefont{Huelga}},
  \bibnamefont{and} \bibinfo{author}{\bibfnamefont{M.~B.}
  \bibnamefont{Plenio}}, \bibinfo{journal}{Physical Review Letters}
  \textbf{\bibinfo{volume}{105}}, \bibinfo{pages}{050404}
  (\bibinfo{year}{2010}), ISSN \bibinfo{issn}{0031-9007},
  \urlprefix\url{http://link.aps.org/doi/10.1103/PhysRevLett.105.050404
  https://link.aps.org/doi/10.1103/PhysRevLett.105.050404}.

\bibitem[{\citenamefont{Suess et~al.}(2014)\citenamefont{Suess, Eisfeld, and
  Strunz}}]{Suess2014}
\bibinfo{author}{\bibfnamefont{D.}~\bibnamefont{Suess}},
  \bibinfo{author}{\bibfnamefont{A.}~\bibnamefont{Eisfeld}}, \bibnamefont{and}
  \bibinfo{author}{\bibfnamefont{W.~T.} \bibnamefont{Strunz}},
  \bibinfo{journal}{Physical Review Letters} \textbf{\bibinfo{volume}{113}},
  \bibinfo{pages}{150403} (\bibinfo{year}{2014}), ISSN
  \bibinfo{issn}{0031-9007}, \eprint{1402.4647},
  \urlprefix\url{https://link.aps.org/doi/10.1103/PhysRevLett.113.150403}.

\bibitem[{\citenamefont{Ol{\v{s}}ina et~al.}(2014)\citenamefont{Ol{\v{s}}ina,
  Kramer, Kreisbeck, and Man{\v{c}}al}}]{Olsina2014}
\bibinfo{author}{\bibfnamefont{J.}~\bibnamefont{Ol{\v{s}}ina}},
  \bibinfo{author}{\bibfnamefont{T.}~\bibnamefont{Kramer}},
  \bibinfo{author}{\bibfnamefont{C.}~\bibnamefont{Kreisbeck}},
  \bibnamefont{and}
  \bibinfo{author}{\bibfnamefont{T.}~\bibnamefont{Man{\v{c}}al}},
  \bibinfo{journal}{The Journal of Chemical Physics}
  \textbf{\bibinfo{volume}{141}}, \bibinfo{pages}{164109}
  (\bibinfo{year}{2014}), ISSN \bibinfo{issn}{0021-9606},
  \urlprefix\url{http://dx.doi.org/10.1063/1.4898354
  http://scitation.aip.org/content/aip/journal/jcp/141/16/10.1063/1.4898354}.

\bibitem[{\citenamefont{Tanimura and Kubo}(1989)}]{Tanimura1989}
\bibinfo{author}{\bibfnamefont{Y.}~\bibnamefont{Tanimura}} \bibnamefont{and}
  \bibinfo{author}{\bibfnamefont{R.}~\bibnamefont{Kubo}},
  \bibinfo{journal}{Journal of the Physics Society Japan}
  \textbf{\bibinfo{volume}{58}}, \bibinfo{pages}{101} (\bibinfo{year}{1989}),
  ISSN \bibinfo{issn}{0031-9015},
  \urlprefix\url{http://jpsj.ipap.jp/link?JPSJ/58/101/}.

\bibitem[{\citenamefont{Tanimura and Mukamel}(1994)}]{Tanimura1994}
\bibinfo{author}{\bibfnamefont{Y.}~\bibnamefont{Tanimura}} \bibnamefont{and}
  \bibinfo{author}{\bibfnamefont{S.}~\bibnamefont{Mukamel}},
  \bibinfo{journal}{Journal of the Physics Society Japan}
  \textbf{\bibinfo{volume}{63}}, \bibinfo{pages}{66} (\bibinfo{year}{1994}).

\bibitem[{\citenamefont{Jin et~al.}(2008)\citenamefont{Jin, Zheng, and
  Yan}}]{Jin2008}
\bibinfo{author}{\bibfnamefont{J.}~\bibnamefont{Jin}},
  \bibinfo{author}{\bibfnamefont{X.}~\bibnamefont{Zheng}}, \bibnamefont{and}
  \bibinfo{author}{\bibfnamefont{Y.}~\bibnamefont{Yan}}, \bibinfo{journal}{The
  Journal of Chemical Physics} \textbf{\bibinfo{volume}{128}},
  \bibinfo{pages}{234703} (\bibinfo{year}{2008}), ISSN
  \bibinfo{issn}{0021-9606}, \eprint{0710.5367},
  \urlprefix\url{http://aip.scitation.org/doi/10.1063/1.2938087}.

\bibitem[{\citenamefont{Ye et~al.}(2016)\citenamefont{Ye, Wang, Hou, Xu, Zheng,
  and Yan}}]{Ye2016}
\bibinfo{author}{\bibfnamefont{L.}~\bibnamefont{Ye}},
  \bibinfo{author}{\bibfnamefont{X.}~\bibnamefont{Wang}},
  \bibinfo{author}{\bibfnamefont{D.}~\bibnamefont{Hou}},
  \bibinfo{author}{\bibfnamefont{R.-X.} \bibnamefont{Xu}},
  \bibinfo{author}{\bibfnamefont{X.}~\bibnamefont{Zheng}}, \bibnamefont{and}
  \bibinfo{author}{\bibfnamefont{Y.}~\bibnamefont{Yan}},
  \bibinfo{journal}{Wiley Interdisciplinary Reviews: Computational Molecular
  Science}  (\bibinfo{year}{2016}), ISSN \bibinfo{issn}{17590876},
  \urlprefix\url{http://doi.wiley.com/10.1002/wcms.1269}.

\bibitem[{\citenamefont{Tsuchimoto and Tanimura}(2015)}]{Tsuchimoto2015}
\bibinfo{author}{\bibfnamefont{M.}~\bibnamefont{Tsuchimoto}} \bibnamefont{and}
  \bibinfo{author}{\bibfnamefont{Y.}~\bibnamefont{Tanimura}},
  \bibinfo{journal}{Journal of Chemical Theory and Computation}
  \textbf{\bibinfo{volume}{11}}, \bibinfo{pages}{3859} (\bibinfo{year}{2015}),
  ISSN \bibinfo{issn}{1549-9618},
  \urlprefix\url{http://pubs.acs.org/doi/abs/10.1021/acs.jctc.5b00488}.

\bibitem[{\citenamefont{Kato and Tanimura}(2016)}]{Kato2016a}
\bibinfo{author}{\bibfnamefont{A.}~\bibnamefont{Kato}} \bibnamefont{and}
  \bibinfo{author}{\bibfnamefont{Y.}~\bibnamefont{Tanimura}},
  \bibinfo{journal}{The Journal of Chemical Physics}
  \textbf{\bibinfo{volume}{145}}, \bibinfo{pages}{224105}
  (\bibinfo{year}{2016}), ISSN \bibinfo{issn}{0021-9606}, \eprint{1609.08783},
  \urlprefix\url{http://aip.scitation.org/doi/10.1063/1.4971370
  http://arxiv.org/abs/1609.08783 http://dx.doi.org/10.1063/1.4971370}.

\bibitem[{\citenamefont{Chen et~al.}(2009)\citenamefont{Chen, Zheng, Shi, and
  Yan}}]{Chen2009}
\bibinfo{author}{\bibfnamefont{L.}~\bibnamefont{Chen}},
  \bibinfo{author}{\bibfnamefont{R.}~\bibnamefont{Zheng}},
  \bibinfo{author}{\bibfnamefont{Q.}~\bibnamefont{Shi}}, \bibnamefont{and}
  \bibinfo{author}{\bibfnamefont{Y.}~\bibnamefont{Yan}},
  \bibinfo{journal}{Journal of Chemical Physics}
  \textbf{\bibinfo{volume}{131}}, \bibinfo{pages}{094502}
  (\bibinfo{year}{2009}), ISSN \bibinfo{issn}{00219606},
  \urlprefix\url{http://www.ncbi.nlm.nih.gov/pubmed/19739856}.

\bibitem[{\citenamefont{Chen et~al.}(2010)\citenamefont{Chen, Zheng, Shi, and
  Yan}}]{Chen2010}
\bibinfo{author}{\bibfnamefont{L.}~\bibnamefont{Chen}},
  \bibinfo{author}{\bibfnamefont{R.}~\bibnamefont{Zheng}},
  \bibinfo{author}{\bibfnamefont{Q.}~\bibnamefont{Shi}}, \bibnamefont{and}
  \bibinfo{author}{\bibfnamefont{Y.}~\bibnamefont{Yan}}, \bibinfo{journal}{The
  Journal of Chemical Physics} \textbf{\bibinfo{volume}{132}},
  \bibinfo{pages}{024505} (\bibinfo{year}{2010}), ISSN
  \bibinfo{issn}{00219606},
  \urlprefix\url{http://www.ncbi.nlm.nih.gov/pubmed/20095685
  http://scitation.aip.org/content/aip/journal/jcp/132/2/10.1063/1.3293039}.

\bibitem[{\citenamefont{Ishizaki and Tanimura}(2008)}]{Ishizaki2008}
\bibinfo{author}{\bibfnamefont{A.}~\bibnamefont{Ishizaki}} \bibnamefont{and}
  \bibinfo{author}{\bibfnamefont{Y.}~\bibnamefont{Tanimura}},
  \bibinfo{journal}{Chemical Physics} \textbf{\bibinfo{volume}{347}},
  \bibinfo{pages}{185} (\bibinfo{year}{2008}), ISSN \bibinfo{issn}{03010104},
  \urlprefix\url{http://linkinghub.elsevier.com/retrieve/pii/S0301010407005150}.

\bibitem[{\citenamefont{Ishizaki and
  Fleming}(2009{\natexlab{b}})}]{Ishizaki2009h}
\bibinfo{author}{\bibfnamefont{A.}~\bibnamefont{Ishizaki}} \bibnamefont{and}
  \bibinfo{author}{\bibfnamefont{G.~R.} \bibnamefont{Fleming}},
  \bibinfo{journal}{Proceedings of the National Academy of Sciences of the
  United States of America} \textbf{\bibinfo{volume}{106}},
  \bibinfo{pages}{17255} (\bibinfo{year}{2009}{\natexlab{b}}), ISSN
  \bibinfo{issn}{0027-8424},
  \urlprefix\url{http://www.annualreviews.org/doi/abs/10.1146/annurev-conmatphys-020911-125126}.

\bibitem[{\citenamefont{Nuernberger et~al.}(2015)\citenamefont{Nuernberger,
  Ruetzel, and Brixner}}]{Nuernberger2015}
\bibinfo{author}{\bibfnamefont{P.}~\bibnamefont{Nuernberger}},
  \bibinfo{author}{\bibfnamefont{S.}~\bibnamefont{Ruetzel}}, \bibnamefont{and}
  \bibinfo{author}{\bibfnamefont{T.}~\bibnamefont{Brixner}},
  \bibinfo{journal}{Angewandte Chemie International Edition}
  \textbf{\bibinfo{volume}{54}}, \bibinfo{pages}{11368} (\bibinfo{year}{2015}),
  ISSN \bibinfo{issn}{14337851},
  \urlprefix\url{http://doi.wiley.com/10.1002/anie.201502974}.

\bibitem[{\citenamefont{Dost{\'{a}}l et~al.}(2016)\citenamefont{Dost{\'{a}}l,
  P{\v{s}}en{\v{c}}{\'{i}}k, and Zigmantas}}]{Dostal2016}
\bibinfo{author}{\bibfnamefont{J.}~\bibnamefont{Dost{\'{a}}l}},
  \bibinfo{author}{\bibfnamefont{J.}~\bibnamefont{P{\v{s}}en{\v{c}}{\'{i}}k}},
  \bibnamefont{and}
  \bibinfo{author}{\bibfnamefont{D.}~\bibnamefont{Zigmantas}},
  \bibinfo{journal}{Nature Chemistry} \textbf{\bibinfo{volume}{8}},
  \bibinfo{pages}{705} (\bibinfo{year}{2016}), ISSN \bibinfo{issn}{1755-4330},
  \urlprefix\url{http://www.nature.com/doifinder/10.1038/nchem.2525}.

\bibitem[{\citenamefont{Blankenship}(2014)}]{Blankenship2014}
\bibinfo{author}{\bibfnamefont{R.~E.} \bibnamefont{Blankenship}},
  \emph{\bibinfo{title}{{Molecular Mechanisms of Photosynthesis}}}
  (\bibinfo{publisher}{Wiley}, \bibinfo{address}{Oxford, UK},
  \bibinfo{year}{2014}), \bibinfo{edition}{2nd} ed.

\bibitem[{\citenamefont{Reimers
  et~al.}(2016{\natexlab{a}})\citenamefont{Reimers, Biczysko, Bruce, Coker,
  Frankcombe, Hashimoto, Hauer, Jankowiak, Kramer, Linnanto
  et~al.}}]{Reimers2016}
\bibinfo{author}{\bibfnamefont{J.~R.} \bibnamefont{Reimers}},
  \bibinfo{author}{\bibfnamefont{M.}~\bibnamefont{Biczysko}},
  \bibinfo{author}{\bibfnamefont{D.}~\bibnamefont{Bruce}},
  \bibinfo{author}{\bibfnamefont{D.~F.} \bibnamefont{Coker}},
  \bibinfo{author}{\bibfnamefont{T.~J.} \bibnamefont{Frankcombe}},
  \bibinfo{author}{\bibfnamefont{H.}~\bibnamefont{Hashimoto}},
  \bibinfo{author}{\bibfnamefont{J.}~\bibnamefont{Hauer}},
  \bibinfo{author}{\bibfnamefont{R.}~\bibnamefont{Jankowiak}},
  \bibinfo{author}{\bibfnamefont{T.}~\bibnamefont{Kramer}},
  \bibinfo{author}{\bibfnamefont{J.}~\bibnamefont{Linnanto}},
  \bibnamefont{et~al.}, \bibinfo{journal}{Biochimica et Biophysica Acta (BBA) -
  Bioenergetics} \textbf{\bibinfo{volume}{1857}}, \bibinfo{pages}{1627}
  (\bibinfo{year}{2016}{\natexlab{a}}), ISSN \bibinfo{issn}{00052728},
  \urlprefix\url{http://linkinghub.elsevier.com/retrieve/pii/S0005272816305709}.

\bibitem[{\citenamefont{Scholes et~al.}(2011)\citenamefont{Scholes, Fleming,
  Olaya-Castro, and {Van Grondelle}}}]{Scholes2011}
\bibinfo{author}{\bibfnamefont{G.~D.} \bibnamefont{Scholes}},
  \bibinfo{author}{\bibfnamefont{G.~R.} \bibnamefont{Fleming}},
  \bibinfo{author}{\bibfnamefont{A.}~\bibnamefont{Olaya-Castro}},
  \bibnamefont{and} \bibinfo{author}{\bibfnamefont{R.}~\bibnamefont{{Van
  Grondelle}}}, \bibinfo{journal}{Nature Chemistry}
  \textbf{\bibinfo{volume}{3}}, \bibinfo{pages}{763} (\bibinfo{year}{2011}),
  ISSN \bibinfo{issn}{17554330},
  \urlprefix\url{http://dx.doi.org/10.1038/nchem.1145}.

\bibitem[{\citenamefont{Romero et~al.}(2017)\citenamefont{Romero,
  Novoderezhkin, and {Van Grondelle}}}]{Romero2017}
\bibinfo{author}{\bibfnamefont{E.}~\bibnamefont{Romero}},
  \bibinfo{author}{\bibfnamefont{V.~I.} \bibnamefont{Novoderezhkin}},
  \bibnamefont{and} \bibinfo{author}{\bibfnamefont{R.}~\bibnamefont{{Van
  Grondelle}}}, \bibinfo{journal}{Nature} \textbf{\bibinfo{volume}{543}},
  \bibinfo{pages}{355} (\bibinfo{year}{2017}), ISSN \bibinfo{issn}{14764687}.

\bibitem[{\citenamefont{Brixner et~al.}(2005)\citenamefont{Brixner, Stenger,
  Vaswani, Cho, Blankenship, and Fleming}}]{Brixner2005}
\bibinfo{author}{\bibfnamefont{T.}~\bibnamefont{Brixner}},
  \bibinfo{author}{\bibfnamefont{J.}~\bibnamefont{Stenger}},
  \bibinfo{author}{\bibfnamefont{H.~M.} \bibnamefont{Vaswani}},
  \bibinfo{author}{\bibfnamefont{M.}~\bibnamefont{Cho}},
  \bibinfo{author}{\bibfnamefont{R.~E.} \bibnamefont{Blankenship}},
  \bibnamefont{and} \bibinfo{author}{\bibfnamefont{G.~R.}
  \bibnamefont{Fleming}}, \bibinfo{journal}{Nature}
  \textbf{\bibinfo{volume}{434}}, \bibinfo{pages}{625} (\bibinfo{year}{2005}),
  ISSN \bibinfo{issn}{0028-0836},
  \urlprefix\url{http://www.ncbi.nlm.nih.gov/pubmed/15800619
  http://www.nature.com/doifinder/10.1038/nature03429}.

\bibitem[{\citenamefont{Engel et~al.}(2007)\citenamefont{Engel, Calhoun, Read,
  Ahn, Man{\v{c}}al, Cheng, Blankenship, and Fleming}}]{Engel2007a}
\bibinfo{author}{\bibfnamefont{G.~S.} \bibnamefont{Engel}},
  \bibinfo{author}{\bibfnamefont{T.~R.} \bibnamefont{Calhoun}},
  \bibinfo{author}{\bibfnamefont{E.~L.} \bibnamefont{Read}},
  \bibinfo{author}{\bibfnamefont{T.-K.} \bibnamefont{Ahn}},
  \bibinfo{author}{\bibfnamefont{T.}~\bibnamefont{Man{\v{c}}al}},
  \bibinfo{author}{\bibfnamefont{Y.-C.} \bibnamefont{Cheng}},
  \bibinfo{author}{\bibfnamefont{R.~E.} \bibnamefont{Blankenship}},
  \bibnamefont{and} \bibinfo{author}{\bibfnamefont{G.~R.}
  \bibnamefont{Fleming}}, \bibinfo{journal}{Nature}
  \textbf{\bibinfo{volume}{446}}, \bibinfo{pages}{782} (\bibinfo{year}{2007}),
  ISSN \bibinfo{issn}{0028-0836},
  \urlprefix\url{http://www.ncbi.nlm.nih.gov/pubmed/17429397
  http://www.nature.com/doifinder/10.1038/nature05678}.

\bibitem[{\citenamefont{May and K{\"{u}}hn}(2004)}]{May2004}
\bibinfo{author}{\bibfnamefont{V.}~\bibnamefont{May}} \bibnamefont{and}
  \bibinfo{author}{\bibfnamefont{O.}~\bibnamefont{K{\"{u}}hn}},
  \emph{\bibinfo{title}{{Charge and Energy Transfer Dynamics in Molecular
  Systems}}} (\bibinfo{publisher}{Wiley-VCH}, \bibinfo{address}{Weinheim},
  \bibinfo{year}{2004}), ISBN \bibinfo{isbn}{3-527-40396-5}.

\bibitem[{\citenamefont{Kreisbeck and Kramer}(2012)}]{Kreisbeck2012b}
\bibinfo{author}{\bibfnamefont{C.}~\bibnamefont{Kreisbeck}} \bibnamefont{and}
  \bibinfo{author}{\bibfnamefont{T.}~\bibnamefont{Kramer}},
  \bibinfo{journal}{Journal of Physical Chemistry Letters}
  \textbf{\bibinfo{volume}{3}}, \bibinfo{pages}{2828} (\bibinfo{year}{2012}),
  ISSN \bibinfo{issn}{19487185}, \eprint{1203.1485}.

\bibitem[{\citenamefont{Kramer and Kreisbeck}(2014)}]{Kramer2014}
\bibinfo{author}{\bibfnamefont{T.}~\bibnamefont{Kramer}} \bibnamefont{and}
  \bibinfo{author}{\bibfnamefont{C.}~\bibnamefont{Kreisbeck}},
  \bibinfo{journal}{AIP Conference Proceedings}
  \textbf{\bibinfo{volume}{1575}}, \bibinfo{pages}{111} (\bibinfo{year}{2014}),
  \urlprefix\url{http://scitation.aip.org/content/aip/proceeding/aipcp/10.1063/1.4861701}.

\bibitem[{\citenamefont{Tanimura}(2006)}]{Tanimura2006}
\bibinfo{author}{\bibfnamefont{Y.}~\bibnamefont{Tanimura}},
  \bibinfo{journal}{Journal of the Physics Society Japan}
  \textbf{\bibinfo{volume}{75}}, \bibinfo{pages}{082001}
  (\bibinfo{year}{2006}), ISSN \bibinfo{issn}{0031-9015},
  \urlprefix\url{http://jpsj.ipap.jp/link?JPSJ/75/082001/}.

\bibitem[{\citenamefont{Hu et~al.}(2010)\citenamefont{Hu, Xu, and
  Yan}}]{Hu2010a}
\bibinfo{author}{\bibfnamefont{J.}~\bibnamefont{Hu}},
  \bibinfo{author}{\bibfnamefont{R.-X.} \bibnamefont{Xu}}, \bibnamefont{and}
  \bibinfo{author}{\bibfnamefont{Y.}~\bibnamefont{Yan}}, \bibinfo{journal}{The
  Journal of Chemical Physics} \textbf{\bibinfo{volume}{133}},
  \bibinfo{pages}{101106} (\bibinfo{year}{2010}), ISSN
  \bibinfo{issn}{0021-9606},
  \urlprefix\url{http://aip.scitation.org/doi/10.1063/1.3484491}.

\bibitem[{\citenamefont{Shi et~al.}(2009)\citenamefont{Shi, Chen, Nan, Xu, and
  Yan}}]{Shi2009b}
\bibinfo{author}{\bibfnamefont{Q.}~\bibnamefont{Shi}},
  \bibinfo{author}{\bibfnamefont{L.}~\bibnamefont{Chen}},
  \bibinfo{author}{\bibfnamefont{G.}~\bibnamefont{Nan}},
  \bibinfo{author}{\bibfnamefont{R.~X.} \bibnamefont{Xu}}, \bibnamefont{and}
  \bibinfo{author}{\bibfnamefont{Y.}~\bibnamefont{Yan}},
  \bibinfo{journal}{Journal of Chemical Physics}
  \textbf{\bibinfo{volume}{130}}, \bibinfo{pages}{084105}
  (\bibinfo{year}{2009}), ISSN \bibinfo{issn}{00219606}.

\bibitem[{\citenamefont{Kreisbeck et~al.}(2014)\citenamefont{Kreisbeck, Kramer,
  and Aspuru-Guzik}}]{Kreisbeck2014}
\bibinfo{author}{\bibfnamefont{C.}~\bibnamefont{Kreisbeck}},
  \bibinfo{author}{\bibfnamefont{T.}~\bibnamefont{Kramer}}, \bibnamefont{and}
  \bibinfo{author}{\bibfnamefont{A.}~\bibnamefont{Aspuru-Guzik}},
  \bibinfo{journal}{Journal of Chemical Theory and Computation}
  \textbf{\bibinfo{volume}{10}}, \bibinfo{pages}{4045} (\bibinfo{year}{2014}),
  ISSN \bibinfo{issn}{1549-9618},
  \urlprefix\url{http://pubs.acs.org/doi/abs/10.1021/ct500629s}.

\bibitem[{\citenamefont{Mascherpa et~al.}(2017)\citenamefont{Mascherpa, Smirne,
  Huelga, and Plenio}}]{Mascherpa2017}
\bibinfo{author}{\bibfnamefont{F.}~\bibnamefont{Mascherpa}},
  \bibinfo{author}{\bibfnamefont{A.}~\bibnamefont{Smirne}},
  \bibinfo{author}{\bibfnamefont{S.~F.} \bibnamefont{Huelga}},
  \bibnamefont{and} \bibinfo{author}{\bibfnamefont{M.~B.}
  \bibnamefont{Plenio}}, \bibinfo{journal}{Physical Review Letters}
  \textbf{\bibinfo{volume}{118}}, \bibinfo{pages}{100401}
  (\bibinfo{year}{2017}), ISSN \bibinfo{issn}{0031-9007},
  \eprint{arXiv:1611.03377v2},
  \urlprefix\url{https://link.aps.org/doi/10.1103/PhysRevLett.118.100401}.

\bibitem[{\citenamefont{Chen et~al.}(2015)\citenamefont{Chen, Gelin, Domcke,
  and Zhao}}]{Chen2015}
\bibinfo{author}{\bibfnamefont{L.}~\bibnamefont{Chen}},
  \bibinfo{author}{\bibfnamefont{M.~F.} \bibnamefont{Gelin}},
  \bibinfo{author}{\bibfnamefont{W.}~\bibnamefont{Domcke}}, \bibnamefont{and}
  \bibinfo{author}{\bibfnamefont{Y.}~\bibnamefont{Zhao}},
  \bibinfo{journal}{Journal of Chemical Physics} \textbf{\bibinfo{volume}{142}}
  (\bibinfo{year}{2015}), ISSN \bibinfo{issn}{00219606},
  \urlprefix\url{http://scitation.aip.org/content/aip/journal/jcp/142/16/10.1063/1.4919240}.

\bibitem[{\citenamefont{Gelin et~al.}(2013)\citenamefont{Gelin, Tanimura, and
  Domcke}}]{Gelin2013}
\bibinfo{author}{\bibfnamefont{M.~F.} \bibnamefont{Gelin}},
  \bibinfo{author}{\bibfnamefont{Y.}~\bibnamefont{Tanimura}}, \bibnamefont{and}
  \bibinfo{author}{\bibfnamefont{W.}~\bibnamefont{Domcke}},
  \bibinfo{journal}{The Journal of Chemical Physics}
  \textbf{\bibinfo{volume}{139}}, \bibinfo{pages}{214302}
  (\bibinfo{year}{2013}), ISSN \bibinfo{issn}{0021-9606},
  \urlprefix\url{http://aip.scitation.org/doi/10.1063/1.4832876}.

\bibitem[{\citenamefont{Hamm and Zanni}(2011)}]{Hamm2011}
\bibinfo{author}{\bibfnamefont{P.}~\bibnamefont{Hamm}} \bibnamefont{and}
  \bibinfo{author}{\bibfnamefont{M.~T.} \bibnamefont{Zanni}},
  \emph{\bibinfo{title}{{Concepts of 2D spectroscopy}}}
  (\bibinfo{publisher}{Cambridge University Press},
  \bibinfo{address}{Cambridge}, \bibinfo{year}{2011}), ISBN
  \bibinfo{isbn}{9781107000056}.

\bibitem[{\citenamefont{Cho et~al.}(2005)\citenamefont{Cho, Vaswani, Brixner,
  Stenger, and Fleming}}]{Cho2005}
\bibinfo{author}{\bibfnamefont{M.}~\bibnamefont{Cho}},
  \bibinfo{author}{\bibfnamefont{H.~M.} \bibnamefont{Vaswani}},
  \bibinfo{author}{\bibfnamefont{T.}~\bibnamefont{Brixner}},
  \bibinfo{author}{\bibfnamefont{J.}~\bibnamefont{Stenger}}, \bibnamefont{and}
  \bibinfo{author}{\bibfnamefont{G.~R.} \bibnamefont{Fleming}},
  \bibinfo{journal}{The Journal of Physical Chemistry B}
  \textbf{\bibinfo{volume}{109}}, \bibinfo{pages}{10542}
  (\bibinfo{year}{2005}), ISSN \bibinfo{issn}{1520-6106},
  \urlprefix\url{http://www.ncbi.nlm.nih.gov/pubmed/16852278
  http://pubs.acs.org/doi/abs/10.1021/jp050788d}.

\bibitem[{\citenamefont{Hein et~al.}(2012)\citenamefont{Hein, Kreisbeck,
  Kramer, and Rodr{\'{i}}guez}}]{Hein2012}
\bibinfo{author}{\bibfnamefont{B.}~\bibnamefont{Hein}},
  \bibinfo{author}{\bibfnamefont{C.}~\bibnamefont{Kreisbeck}},
  \bibinfo{author}{\bibfnamefont{T.}~\bibnamefont{Kramer}}, \bibnamefont{and}
  \bibinfo{author}{\bibfnamefont{M.}~\bibnamefont{Rodr{\'{i}}guez}},
  \bibinfo{journal}{New Journal of Physics} \textbf{\bibinfo{volume}{14}},
  \bibinfo{pages}{023018} (\bibinfo{year}{2012}), ISSN
  \bibinfo{issn}{13672630}, \eprint{1110.1511},
  \urlprefix\url{http://stacks.iop.org/1367-2630/14/i=2/a=023018?key=crossref.6311869885668512ae4c7fbf043da01c}.

\bibitem[{\citenamefont{Gelin et~al.}(2017)\citenamefont{Gelin, Borrelli, and
  Domcke}}]{Gelin2017}
\bibinfo{author}{\bibfnamefont{M.~F.} \bibnamefont{Gelin}},
  \bibinfo{author}{\bibfnamefont{R.}~\bibnamefont{Borrelli}}, \bibnamefont{and}
  \bibinfo{author}{\bibfnamefont{W.}~\bibnamefont{Domcke}},
  \bibinfo{journal}{The Journal of Chemical Physics}
  \textbf{\bibinfo{volume}{147}}, \bibinfo{pages}{044114}
  (\bibinfo{year}{2017}), ISSN \bibinfo{issn}{0021-9606},
  \urlprefix\url{http://aip.scitation.org/doi/10.1063/1.4996205}.

\bibitem[{\citenamefont{Gordon}(1968)}]{Gordon1968a}
\bibinfo{author}{\bibfnamefont{R.}~\bibnamefont{Gordon}}, in
  \emph{\bibinfo{booktitle}{Advances in Magnetic and Optical Resonance}}
  (\bibinfo{publisher}{ACADEMIC PRESS INC.}, \bibinfo{year}{1968}),
  vol.~\bibinfo{volume}{3}, pp. \bibinfo{pages}{1--42},
  \urlprefix\url{http://dx.doi.org/10.1016/B978-1-4832-3116-7.50008-4
  http://linkinghub.elsevier.com/retrieve/pii/B9781483231167500084}.

\bibitem[{\citenamefont{Shuang et~al.}(2001)\citenamefont{Shuang, Yang, and
  Yan}}]{Shuang2001}
\bibinfo{author}{\bibfnamefont{F.}~\bibnamefont{Shuang}},
  \bibinfo{author}{\bibfnamefont{C.}~\bibnamefont{Yang}}, \bibnamefont{and}
  \bibinfo{author}{\bibfnamefont{Y.}~\bibnamefont{Yan}}, \bibinfo{journal}{The
  Journal of Chemical Physics} \textbf{\bibinfo{volume}{114}},
  \bibinfo{pages}{3868} (\bibinfo{year}{2001}), ISSN \bibinfo{issn}{00219606},
  \urlprefix\url{http://scitation.aip.org/content/aip/journal/jcp/114/9/10.1063/1.1344608}.

\bibitem[{\citenamefont{Zhang et~al.}(2017)\citenamefont{Zhang, Qiao, Xu,
  Zheng, and Yan}}]{Zhang2017}
\bibinfo{author}{\bibfnamefont{H.-D.} \bibnamefont{Zhang}},
  \bibinfo{author}{\bibfnamefont{Q.}~\bibnamefont{Qiao}},
  \bibinfo{author}{\bibfnamefont{R.-X.} \bibnamefont{Xu}},
  \bibinfo{author}{\bibfnamefont{X.}~\bibnamefont{Zheng}}, \bibnamefont{and}
  \bibinfo{author}{\bibfnamefont{Y.}~\bibnamefont{Yan}}, \bibinfo{journal}{The
  Journal of Chemical Physics} \textbf{\bibinfo{volume}{147}},
  \bibinfo{pages}{044105} (\bibinfo{year}{2017}), ISSN
  \bibinfo{issn}{0021-9606}.

\bibitem[{\citenamefont{Dijkstra and Tanimura}(2010)}]{Dijkstra2010}
\bibinfo{author}{\bibfnamefont{A.~G.} \bibnamefont{Dijkstra}} \bibnamefont{and}
  \bibinfo{author}{\bibfnamefont{Y.}~\bibnamefont{Tanimura}},
  \bibinfo{journal}{Physical Review Letters} \textbf{\bibinfo{volume}{104}},
  \bibinfo{pages}{250401} (\bibinfo{year}{2010}), ISSN
  \bibinfo{issn}{0031-9007},
  \urlprefix\url{http://link.aps.org/doi/10.1103/PhysRevLett.104.250401
  https://link.aps.org/doi/10.1103/PhysRevLett.104.250401}.

\bibitem[{\citenamefont{Seidner et~al.}(1995)\citenamefont{Seidner, Stock, and
  Domcke}}]{Seidner1995}
\bibinfo{author}{\bibfnamefont{L.}~\bibnamefont{Seidner}},
  \bibinfo{author}{\bibfnamefont{G.}~\bibnamefont{Stock}}, \bibnamefont{and}
  \bibinfo{author}{\bibfnamefont{W.}~\bibnamefont{Domcke}},
  \bibinfo{journal}{The Journal of Chemical Physics}
  \textbf{\bibinfo{volume}{103}}, \bibinfo{pages}{3998} (\bibinfo{year}{1995}),
  ISSN \bibinfo{issn}{00219606},
  \urlprefix\url{http://scitation.aip.org/content/aip/journal/jcp/103/10/10.1063/1.469586}.

\bibitem[{\citenamefont{Wolfseder et~al.}(1997)\citenamefont{Wolfseder,
  Seidner, Stock, and Domcke}}]{Wolfseder1997}
\bibinfo{author}{\bibfnamefont{B.}~\bibnamefont{Wolfseder}},
  \bibinfo{author}{\bibfnamefont{L.}~\bibnamefont{Seidner}},
  \bibinfo{author}{\bibfnamefont{G.}~\bibnamefont{Stock}}, \bibnamefont{and}
  \bibinfo{author}{\bibfnamefont{W.}~\bibnamefont{Domcke}},
  \bibinfo{journal}{Chemical Physics} \textbf{\bibinfo{volume}{217}},
  \bibinfo{pages}{275} (\bibinfo{year}{1997}), ISSN \bibinfo{issn}{03010104},
  \urlprefix\url{http://linkinghub.elsevier.com/retrieve/pii/S0301010497000463}.

\bibitem[{\citenamefont{Kramer et~al.}(2017)\citenamefont{Kramer,
  Rodr{\'{i}}guez, and Zelinskyy}}]{Kramer2017a}
\bibinfo{author}{\bibfnamefont{T.}~\bibnamefont{Kramer}},
  \bibinfo{author}{\bibfnamefont{M.}~\bibnamefont{Rodr{\'{i}}guez}},
  \bibnamefont{and}
  \bibinfo{author}{\bibfnamefont{Y.}~\bibnamefont{Zelinskyy}},
  \bibinfo{journal}{The Journal of Physical Chemistry B}
  \textbf{\bibinfo{volume}{121}}, \bibinfo{pages}{463} (\bibinfo{year}{2017}),
  ISSN \bibinfo{issn}{1520-6106},
  \urlprefix\url{http://pubs.acs.org/doi/abs/10.1021/acs.jpcb.6b09858}.

\bibitem[{\citenamefont{Mukamel}(1995)}]{Mukamel1995}
\bibinfo{author}{\bibfnamefont{S.}~\bibnamefont{Mukamel}},
  \emph{\bibinfo{title}{{Principles of Nonlinear Optical Spectroscopy}}}
  (\bibinfo{publisher}{Oxford University Press}, \bibinfo{address}{Oxford},
  \bibinfo{year}{1995}).

\bibitem[{\citenamefont{Chen et~al.}(2011)\citenamefont{Chen, Zheng, Jing, and
  Shi}}]{Chen2011}
\bibinfo{author}{\bibfnamefont{L.}~\bibnamefont{Chen}},
  \bibinfo{author}{\bibfnamefont{R.}~\bibnamefont{Zheng}},
  \bibinfo{author}{\bibfnamefont{Y.}~\bibnamefont{Jing}}, \bibnamefont{and}
  \bibinfo{author}{\bibfnamefont{Q.}~\bibnamefont{Shi}}, \bibinfo{journal}{The
  Journal of Chemical Physics} \textbf{\bibinfo{volume}{134}},
  \bibinfo{pages}{194508} (\bibinfo{year}{2011}), ISSN
  \bibinfo{issn}{00219606},
  \urlprefix\url{http://www.ncbi.nlm.nih.gov/pubmed/21599074
  http://scitation.aip.org/content/aip/journal/jcp/134/19/10.1063/1.3589982}.

\bibitem[{\citenamefont{Craig and Thirunamachandran}(1998)}]{Craig1984}
\bibinfo{author}{\bibfnamefont{D.~P.} \bibnamefont{Craig}} \bibnamefont{and}
  \bibinfo{author}{\bibfnamefont{T.}~\bibnamefont{Thirunamachandran}},
  \emph{\bibinfo{title}{{Molecular Quantum Electrodynamics: An Introduction to
  Radiation-molecule Interactions}}} (\bibinfo{publisher}{Dover Publications},
  \bibinfo{address}{Mineola, New York}, \bibinfo{year}{1998}), ISBN
  \bibinfo{isbn}{9780486402147}.

\bibitem[{\citenamefont{Reimers
  et~al.}(2016{\natexlab{b}})\citenamefont{Reimers, Biczysko, Bruce, Coker,
  Frankcombe, Hashimoto, Hauer, Jankowiak, Kramer, Linnanto
  et~al.}}]{Reimers2016a}
\bibinfo{author}{\bibfnamefont{J.~R.} \bibnamefont{Reimers}},
  \bibinfo{author}{\bibfnamefont{M.}~\bibnamefont{Biczysko}},
  \bibinfo{author}{\bibfnamefont{D.}~\bibnamefont{Bruce}},
  \bibinfo{author}{\bibfnamefont{D.~F.} \bibnamefont{Coker}},
  \bibinfo{author}{\bibfnamefont{T.~J.} \bibnamefont{Frankcombe}},
  \bibinfo{author}{\bibfnamefont{H.}~\bibnamefont{Hashimoto}},
  \bibinfo{author}{\bibfnamefont{J.}~\bibnamefont{Hauer}},
  \bibinfo{author}{\bibfnamefont{R.}~\bibnamefont{Jankowiak}},
  \bibinfo{author}{\bibfnamefont{T.}~\bibnamefont{Kramer}},
  \bibinfo{author}{\bibfnamefont{J.}~\bibnamefont{Linnanto}},
  \bibnamefont{et~al.}, \bibinfo{journal}{Biochimica et Biophysica Acta (BBA) -
  Bioenergetics}  (\bibinfo{year}{2016}{\natexlab{b}}), ISSN
  \bibinfo{issn}{00052728},
  \urlprefix\url{http://linkinghub.elsevier.com/retrieve/pii/S0005272816305709}.

\bibitem[{\citenamefont{Adolphs and Renger}(2006)}]{Adolphs2006a}
\bibinfo{author}{\bibfnamefont{J.}~\bibnamefont{Adolphs}} \bibnamefont{and}
  \bibinfo{author}{\bibfnamefont{T.}~\bibnamefont{Renger}},
  \bibinfo{journal}{Biophysical Journal} \textbf{\bibinfo{volume}{91}},
  \bibinfo{pages}{2778} (\bibinfo{year}{2006}), ISSN \bibinfo{issn}{00063495},
  \urlprefix\url{http://www.ncbi.nlm.nih.gov/pubmed/16861264
  http://linkinghub.elsevier.com/retrieve/pii/S0006349506719932}.

\bibitem[{\citenamefont{Vulto et~al.}(1998)\citenamefont{Vulto, de~Baat, Louwe,
  Permentier, Neef, Miller, van Amerongen, and Aartsma}}]{Vulto1998}
\bibinfo{author}{\bibfnamefont{S.~I.~E.} \bibnamefont{Vulto}},
  \bibinfo{author}{\bibfnamefont{M.~A.} \bibnamefont{de~Baat}},
  \bibinfo{author}{\bibfnamefont{R.~J.~W.} \bibnamefont{Louwe}},
  \bibinfo{author}{\bibfnamefont{H.~P.} \bibnamefont{Permentier}},
  \bibinfo{author}{\bibfnamefont{T.}~\bibnamefont{Neef}},
  \bibinfo{author}{\bibfnamefont{M.}~\bibnamefont{Miller}},
  \bibinfo{author}{\bibfnamefont{H.}~\bibnamefont{van Amerongen}},
  \bibnamefont{and} \bibinfo{author}{\bibfnamefont{T.~J.}
  \bibnamefont{Aartsma}}, \bibinfo{journal}{The Journal of Physical Chemistry
  B} \textbf{\bibinfo{volume}{102}}, \bibinfo{pages}{9577}
  (\bibinfo{year}{1998}), ISSN \bibinfo{issn}{1520-6106},
  \urlprefix\url{http://pubs.acs.org/doi/abs/10.1021/jp982095l}.

\bibitem[{\citenamefont{Davis et~al.}(1998)\citenamefont{Davis, Wasielewski,
  Kosloff, and Ratner}}]{Davis1998a}
\bibinfo{author}{\bibfnamefont{W.~B.} \bibnamefont{Davis}},
  \bibinfo{author}{\bibfnamefont{M.~R.} \bibnamefont{Wasielewski}},
  \bibinfo{author}{\bibfnamefont{R.}~\bibnamefont{Kosloff}}, \bibnamefont{and}
  \bibinfo{author}{\bibfnamefont{M.~A.} \bibnamefont{Ratner}},
  \bibinfo{journal}{The Journal of Physical Chemistry A}
  \textbf{\bibinfo{volume}{102}}, \bibinfo{pages}{9360} (\bibinfo{year}{1998}),
  ISSN \bibinfo{issn}{1089-5639},
  \urlprefix\url{http://pubs.acs.org/doi/abs/10.1021/jp9813544}.

\bibitem[{\citenamefont{Cheng and Silbey}(2005)}]{Cheng2005}
\bibinfo{author}{\bibfnamefont{Y.~C.} \bibnamefont{Cheng}} \bibnamefont{and}
  \bibinfo{author}{\bibfnamefont{R.~J.} \bibnamefont{Silbey}},
  \bibinfo{journal}{Journal of Physical Chemistry B}
  \textbf{\bibinfo{volume}{109}}, \bibinfo{pages}{21399}
  (\bibinfo{year}{2005}), ISSN \bibinfo{issn}{15206106}.

\bibitem[{\citenamefont{Vulto et~al.}(1999)\citenamefont{Vulto, de~Baat,
  Neerken, Nowak, van Amerongen, Amesz, and Aartsma}}]{Vulto1999}
\bibinfo{author}{\bibfnamefont{S.~I.~E.} \bibnamefont{Vulto}},
  \bibinfo{author}{\bibfnamefont{M.~A.} \bibnamefont{de~Baat}},
  \bibinfo{author}{\bibfnamefont{S.}~\bibnamefont{Neerken}},
  \bibinfo{author}{\bibfnamefont{F.~R.} \bibnamefont{Nowak}},
  \bibinfo{author}{\bibfnamefont{H.}~\bibnamefont{van Amerongen}},
  \bibinfo{author}{\bibfnamefont{J.}~\bibnamefont{Amesz}}, \bibnamefont{and}
  \bibinfo{author}{\bibfnamefont{T.~J.} \bibnamefont{Aartsma}},
  \bibinfo{journal}{The Journal of Physical Chemistry B}
  \textbf{\bibinfo{volume}{103}}, \bibinfo{pages}{8153} (\bibinfo{year}{1999}),
  ISSN \bibinfo{issn}{1520-6106},
  \urlprefix\url{http://pubs.acs.org/doi/abs/10.1021/jp984702a}.

\bibitem[{\citenamefont{Adolphs et~al.}(2008)\citenamefont{Adolphs, M{\"{u}}h,
  Madjet, and Renger}}]{Adolphs2008}
\bibinfo{author}{\bibfnamefont{J.}~\bibnamefont{Adolphs}},
  \bibinfo{author}{\bibfnamefont{F.}~\bibnamefont{M{\"{u}}h}},
  \bibinfo{author}{\bibfnamefont{M.~E.-A.} \bibnamefont{Madjet}},
  \bibnamefont{and} \bibinfo{author}{\bibfnamefont{T.}~\bibnamefont{Renger}},
  \bibinfo{journal}{Photosynthesis Research} \textbf{\bibinfo{volume}{95}},
  \bibinfo{pages}{197} (\bibinfo{year}{2008}), ISSN \bibinfo{issn}{0166-8595},
  \urlprefix\url{http://link.springer.com/10.1007/s11120-007-9248-z}.

\bibitem[{\citenamefont{Kramer and Rodriguez}(2017)}]{Kramer2017b}
\bibinfo{author}{\bibfnamefont{T.}~\bibnamefont{Kramer}} \bibnamefont{and}
  \bibinfo{author}{\bibfnamefont{M.}~\bibnamefont{Rodriguez}},
  \bibinfo{journal}{Scientific Reports} \textbf{\bibinfo{volume}{7}},
  \bibinfo{pages}{45245} (\bibinfo{year}{2017}), ISSN
  \bibinfo{issn}{2045-2322},
  \urlprefix\url{http://www.nature.com/articles/srep45245}.

\bibitem[{\citenamefont{Kreisbeck et~al.}(2013)\citenamefont{Kreisbeck, Kramer,
  and Aspuru-Guzik}}]{Kreisbeck2013}
\bibinfo{author}{\bibfnamefont{C.}~\bibnamefont{Kreisbeck}},
  \bibinfo{author}{\bibfnamefont{T.}~\bibnamefont{Kramer}}, \bibnamefont{and}
  \bibinfo{author}{\bibfnamefont{A.}~\bibnamefont{Aspuru-Guzik}},
  \bibinfo{journal}{Journal of Physical Chemistry B}
  \textbf{\bibinfo{volume}{117}}, \bibinfo{pages}{9380} (\bibinfo{year}{2013}),
  ISSN \bibinfo{issn}{15206106}, \eprint{1306.4942}.

\bibitem[{\citenamefont{Thyrhaug et~al.}(2016)\citenamefont{Thyrhaug,
  {\v{Z}}{\'{i}}dek, Dost{\'{a}}l, B{\'{i}}na, and Zigmantas}}]{Thyrhaug2016a}
\bibinfo{author}{\bibfnamefont{E.}~\bibnamefont{Thyrhaug}},
  \bibinfo{author}{\bibfnamefont{K.}~\bibnamefont{{\v{Z}}{\'{i}}dek}},
  \bibinfo{author}{\bibfnamefont{J.}~\bibnamefont{Dost{\'{a}}l}},
  \bibinfo{author}{\bibfnamefont{D.}~\bibnamefont{B{\'{i}}na}},
  \bibnamefont{and}
  \bibinfo{author}{\bibfnamefont{D.}~\bibnamefont{Zigmantas}},
  \bibinfo{journal}{The Journal of Physical Chemistry Letters}
  \textbf{\bibinfo{volume}{7}}, \bibinfo{pages}{1653} (\bibinfo{year}{2016}),
  ISSN \bibinfo{issn}{1948-7185},
  \urlprefix\url{http://pubs.acs.org/doi/abs/10.1021/acs.jpclett.6b00534}.

\bibitem[{\citenamefont{Str{\"{u}}mpfer and Schulten}(2012)}]{Strumpfer2012a}
\bibinfo{author}{\bibfnamefont{J.}~\bibnamefont{Str{\"{u}}mpfer}}
  \bibnamefont{and} \bibinfo{author}{\bibfnamefont{K.}~\bibnamefont{Schulten}},
  \bibinfo{journal}{Journal of Chemical Theory and Computation}
  \textbf{\bibinfo{volume}{8}}, \bibinfo{pages}{2808} (\bibinfo{year}{2012}),
  ISSN \bibinfo{issn}{15499618}.

\bibitem[{\citenamefont{Kreisbeck et~al.}(2011)\citenamefont{Kreisbeck, Kramer,
  Rodr{\'{i}}guez, and Hein}}]{Kreisbeck2011}
\bibinfo{author}{\bibfnamefont{C.}~\bibnamefont{Kreisbeck}},
  \bibinfo{author}{\bibfnamefont{T.}~\bibnamefont{Kramer}},
  \bibinfo{author}{\bibfnamefont{M.}~\bibnamefont{Rodr{\'{i}}guez}},
  \bibnamefont{and} \bibinfo{author}{\bibfnamefont{B.}~\bibnamefont{Hein}},
  \bibinfo{journal}{Journal of Chemical Theory and Computation}
  \textbf{\bibinfo{volume}{7}}, \bibinfo{pages}{2166} (\bibinfo{year}{2011}),
  ISSN \bibinfo{issn}{15499618}, \eprint{1012.4382},
  \urlprefix\url{http://pubs.acs.org/doi/abs/10.1021/ct200126d}.

\bibitem[{\citenamefont{Williams et~al.}(2009)\citenamefont{Williams, Waterman,
  and Patterson}}]{Williams2009}
\bibinfo{author}{\bibfnamefont{S.}~\bibnamefont{Williams}},
  \bibinfo{author}{\bibfnamefont{A.}~\bibnamefont{Waterman}}, \bibnamefont{and}
  \bibinfo{author}{\bibfnamefont{D.}~\bibnamefont{Patterson}},
  \bibinfo{journal}{Communications of the ACM} \textbf{\bibinfo{volume}{52}},
  \bibinfo{pages}{65} (\bibinfo{year}{2009}), ISSN \bibinfo{issn}{00010782},
  \eprint{1103.4300v1}.

\bibitem[{\citenamefont{Noack et~al.}(2018)\citenamefont{Noack, Reinefeld,
  Kramer, and Steinke}}]{Noack2018}
\bibinfo{author}{\bibfnamefont{M.}~\bibnamefont{Noack}},
  \bibinfo{author}{\bibfnamefont{A.}~\bibnamefont{Reinefeld}},
  \bibinfo{author}{\bibfnamefont{T.}~\bibnamefont{Kramer}}, \bibnamefont{and}
  \bibinfo{author}{\bibfnamefont{T.}~\bibnamefont{Steinke}}, in
  \emph{\bibinfo{booktitle}{19th IEEE International Workshop on Parallel and
  Distributed Scientific and Engineering Computing (PDSEC 2018)}}
  (\bibinfo{year}{2018}).

\bibitem[{\citenamefont{Kreisbeck and Kramer}(2014)}]{GPUHEOM}
\bibinfo{author}{\bibfnamefont{C.}~\bibnamefont{Kreisbeck}} \bibnamefont{and}
  \bibinfo{author}{\bibfnamefont{T.}~\bibnamefont{Kramer}},
  \emph{\bibinfo{title}{{Exciton Dynamics Lab for Light-Harvesting Complexes
  (GPU-HEOM)}}} (\bibinfo{year}{2014}),
  \urlprefix\url{dx.doi.org/10.4231/D3RF5KH7G}.

\end{thebibliography}
\end{document}